\documentclass[conference]{IEEEtran}
\usepackage{cite}
\usepackage{amsmath,amssymb,amsfonts}
\usepackage{booktabs}
\usepackage{multirow}
\usepackage{algorithmic}
\usepackage{graphicx}
\usepackage{textcomp}
\usepackage{xcolor}
\usepackage{hyperref}
\usepackage{float}
\usepackage{booktabs}
\usepackage{tabularx}
\hypersetup{
    colorlinks=true,
    linkcolor=blue,
    filecolor=magenta,      
    urlcolor=cyan,
}
\usepackage{graphicx}
\usepackage{array}
\usepackage{tabularx}  % Enables auto-width columns
\usepackage{adjustbox}
\usepackage[margin=1in]{geometry}
\begin{document}

\title{A Comprehensive Energy Management Application Method considering Smart Home Occupant Behavior using IoT and Real Big Data}

\author{
    \IEEEauthorblockN{S. Saba Rafiei, Mahdi S. Naderi, Mehrdad Abedi}
    \IEEEauthorblockA{
        Department of Electrical Engineering, Amirkabir University of Technology, Tehran, Iran \\
        Iran Grid Secure Operation Research Center, Amirkabir University of Technology, Tehran, Iran \\
        Email: saba\_rafiei@aut.ac.ir, salaynaderi@aut.ac.ir, abedi@aut.ac.ir
    }
}

\maketitle

\begin{abstract}
One of the most far-reaching use cases of the internet of things is in smart grid and smart home operation. The smart home concept allows residents to control, monitor, and manage their energy consumption with minimum loss and self-involvement. Since each household's lifestyle and energy consumption is unique, the management system needs background knowledge about residents' energy consumption behavioral patterns for more accurate planning. To obtain this information, data related to residents' consumption records must be processed. This research has attempted to provide an optimal decentralized management system consisting of interoperable sections for forecasting, optimizing, scheduling, and implementing load management on a smart home. Comparing different prediction models using 4 years of 1-min interval real data of a smart home with photovoltaic generation (PV) and electric vehicle (EV), forecasting non-controllable loads and taking a deterministic approach in different scenarios, the system uses mixed integer linear programming (MILP) to provide load scheduling with the objective of an optimal total energy cost reduction with minimum changes in the household's desired consumption compared to the initial state. The results have shown that the proposed system has reliable performance due to the high precision of the forecast and has led to increased energy efficiency, reduced energy cost (up to 62. 05\%), reduced peak-to-average ratio (PAR) (up to 44. 19\%) and reduced standard deviation (SD) (up to 19. 70\%) in net consumption.
\end{abstract}


\section{Introduction}
With the development of artificial intelligence and the increasing complexity of loads, several deep learning methods have been applied in short-term forecasting in recent years. Most of these methods use centralized algorithms to build short-term prediction models. However, with the integration of renewable energies and the rapid development of smart meters, these centralized algorithms face major problems. Dealing with such large-dimensional load data has become more challenging in terms of timely requirements in short-term forecasting, and therefore many researchers have studied distributed frameworks to accelerate the short-term forecasting process. According to \cite{1}, for the proper distribution of power between small portable devices, the integration of the massive Internet of Things system with artificial intelligence-based techniques is an important issue. Researchers in \cite{2} separated short-term forecasting into several local forecasting models and proposed a distributed short-term forecasting model based on local weather information that can increase the accuracy of the system-level load forecasting model; However, this study has only considered temperature and humidity among effective weather conditions with the THI (Temperature-Humidity Index) and uses subsets with similar weather conditions, with 15-minute intervals of recorded data, which lowers the prediction accuracy compared to the work of this paper, in which we use precise geographic coordinates and a more minor time step. A multi-objective mixed integer nonlinear programming model is built using a meaningful balance between energy saving and comfortable lifestyle for the optimal use of energy in a smart home, which is referred to in \cite{3}. This reference considers user comfort to be mainly dependent on three environmental factors, including temperature, relative humidity, and air movement, but only studies temperature. Also, although wind speed is another important factor that increases the heat transfer between the building and the external environment by increasing the penetration and convection heat transfer coefficient, its effect is ignored in this article. Using the prediction model of environmental indicators in smart houses, \cite{4} performs the management of residents' comfort based on energy optimization. Heater electricity consumption is the only data that is used from the objective function to achieve optimal comfort. This study considers only temperature and humidity for the convenience of the user. In \cite{5}, a web-based decision support system has been developed using multi-purpose data in the context of a smart city. The focus of this work was on heating and cooling devices and did not pay particular attention to other electrical appliances. According to \cite{6}, which provides an overview of all the work done in order to optimize energy consumption until 2019, most of the optimizations focused on thermal satisfaction of users and did not pay attention to other effective factors. Also, authors in \cite{7} provided an optimization framework for planning various smart home appliances and energy resource allocation. However, they did not consider the significant role of artificial intelligence in a day-ahead energy plan for the appliances in a smart home. Authors in \cite{8} study a consumer house with considered load response aggregator and an objective function to minimizes daily energy cost, peak-to-average ratio and waiting time delay. However, they assumed uncontrollable loads to be predetermined.

A framework for energy measurement and real-time excitation based on Internet of Things is presented in reference \cite{9}. This framework uses a private cloud setup, managed by the user and not by a third party, by labview software. Since all smart devices can be connected to the Internet directly or through their hub, reference \cite{10} suggests that machine-to-machine communication be done through the cloud. A building management system capable of efficient and automatic management of building elements using human behavioral models is presented in reference \cite{11}. This reference did not use real data for modeling.

Data processing is one of the main steps in energy control of smart homes. More data helps forecasting and management systems operate more accurately. Still, challenges such as heavy computational load, high data volume, bandwidth limitations, and security issues have prevented highly accumulative data from being sent to the processing and computing center. References \cite{12}, \cite{13} and \cite{14} have tried to bring the edge computing resources as close as possible to the data generation reference. In reference \cite{15}, various techniques of compression and optimization of data storage in the cloud, their consequences and future paths are discussed. In an intelligent distributed management system, the calculations are performed on the distributed processing layers of the network. With the investigations carried out in \cite{1}, an interlayer approach provides more energy efficiency and reliable and strong performance for the joint application of artificial intelligence and the Internet of Things. Y sun et. al in \cite{16} introduce IntelliHome as a smart home system that aims to reduce electricity consumption in the household. However, consumption forecast is done based on a 10-month synthetic dataset, leading to low accuracy in results. In this study, we use big real data related to 4 consecutive years (2054880 minutes) to make accurate predictions, get more reliable results with real-world information and avoid assumptions far from reality.
With the integration of renewable energy resources, battery energy storage systems, and controllable loads, the power grid becomes distributed and complex. The operation status of the power grid may change frequently, so the reasonable energy management strategy of the power grid is a critical aspect in smart grid research, which is designed to meet consumer demand in different time intervals and realize the efficient operation of the smart grid. 

Due to the intermittent nature and unpredictability of renewable energies, the future smart grid will inevitably combine more dynamic elements. Meanwhile, the smart grid must maintain a balance between supply and demand. Renewable energy sources can have considerable fluctuations in output, even at intervals of one minute or less. In the traditional grid operating model, distributed generation (usually fossil fuel generators) are tuned by increasing or decreasing their outputs to ensure that demand is met \cite{17}. 

In an IoT scenario, the variability of renewable energy production can be measured directly by IoT devices on the grid, sensors installed on the grid, or decentralized control. IoT devices can react quickly by increasing or decreasing electrical loads and reducing the harmful effects of variations in renewable energy production on the grid.

In the proposed model of this study, the information is collected locally in real-time, and the predictions and calculations of the smart home planning are done on-site as a nano-grid with the ability to communicate with the control center and the network through IoT gateways. 

Since accurate and correct load forecasting has a definitive role in energy management, comparison in the performance of consumption prediction models through data-driven prediction using Recurrent Neural Networks (RNNs), Long-Short Term Memory (LSTM) networks, Bidirectional Long-Short Term Memory (BLSTM) networks, Gated Recurrent Unit (GRU), Bidirectional Gated Recurrent Unit (BGRU), Stacked LSTM and GUR (SLSTM \& SGRU), Hybrid RNN, and also sequence-to-sequence (Seq2Seq) model has been done \cite{18, 19, 20, 21, 22}. 

In previous studies, the load prediction was mainly made on controllable or thermal loads \cite{23},~\cite{24}. Also, load scheduling was done without accurate consideration of non-controllable loads \cite{25}. None of these studies opted to show the non-negligible effects of non-controllable loads on energy management programs in a smart home. 

In this study, the aim is also to predict non-controllable loads with maximum accuracy, build the day-ahead load-scheduling plan with consideration of these loads, and investigate the effects of these loads on the optimal energy consumption plan and the total consumption profile for the next day.

The rest of the paper is organized as follows: Section~\ref{sec:2} discusses the structure of the management system consisting of different units and presents an overall understanding of the role that each unit plays in the process, along with their modeling, configuration, and formulations. 

Section~\ref{sec:3} explains the scenarios and assumptions used to validate the proposed energy management system. 

Sections \ref{sec:4} and \ref{sec:5} include different results analysis and conclusions of the present work.

\section{Structure of the Energy Management System}
\label{sec:2}
The architecture of the proposed system model includes power transmission and communication by which smart appliances and plugs can be controlled. The control unit is considered as the computing center of the management system. In addition, the ability to communicate, record, and monitor raw data as well as engineered data, enables the extraction of information related to consumption and production profiles. Appliance programming is performed by the control unit software, which includes a built-in optimization solver. This system uses a web crawler to get electricity price information and weather information from reliable sources. Also, home consumption information, recorded by smart sensors, is sent to the database and stored and processed in the fog through the Internet of Things (IoT) gateways. The information needed for more complex processing and calculations is transferred to the cloud, thus enabling a two-way exchange of information with the network operator. This system can receive and send electricity price data and energy consumption statistics for purchase or sale. The forecasting unit is set to choose the best prediction model structure according to the received data and provide the system with accurate day-ahead consumption forecasts. Constrained optimization is done to bring the home consumption profile closer to instantaneous renewable production and reduce costs by considering the residents' thermal comfort and preferences. The results of the optimal day-ahead scheduling are converted into decisions and sent to the actuators installed in smart sockets or smart appliances equipped with IoT modules with specific Internet Protocol addresses (IP addresses). The sensors record the appliance's power consumption every minute and send the information to the database. This information will again be sent to the forecasting unit for processing to increase the forecast's accuracy and will play its role in planning the next day's consumption. If a new device or consumer unit is added to the home, the system detects it. After IP assignment, the consumption data is updated, and future predictions, calculations, and decisions are adapted. It should also be noted that to prevent the saturation of the database capacity with the addition of new data, after processing the data once in the prediction section and obtaining the weight coefficients of the desired network, the previous data is dropped from the database, and the weight coefficients are updated every day with new data.

\begin{figure*}[t]
    \centering
    \includegraphics[width=\textwidth]{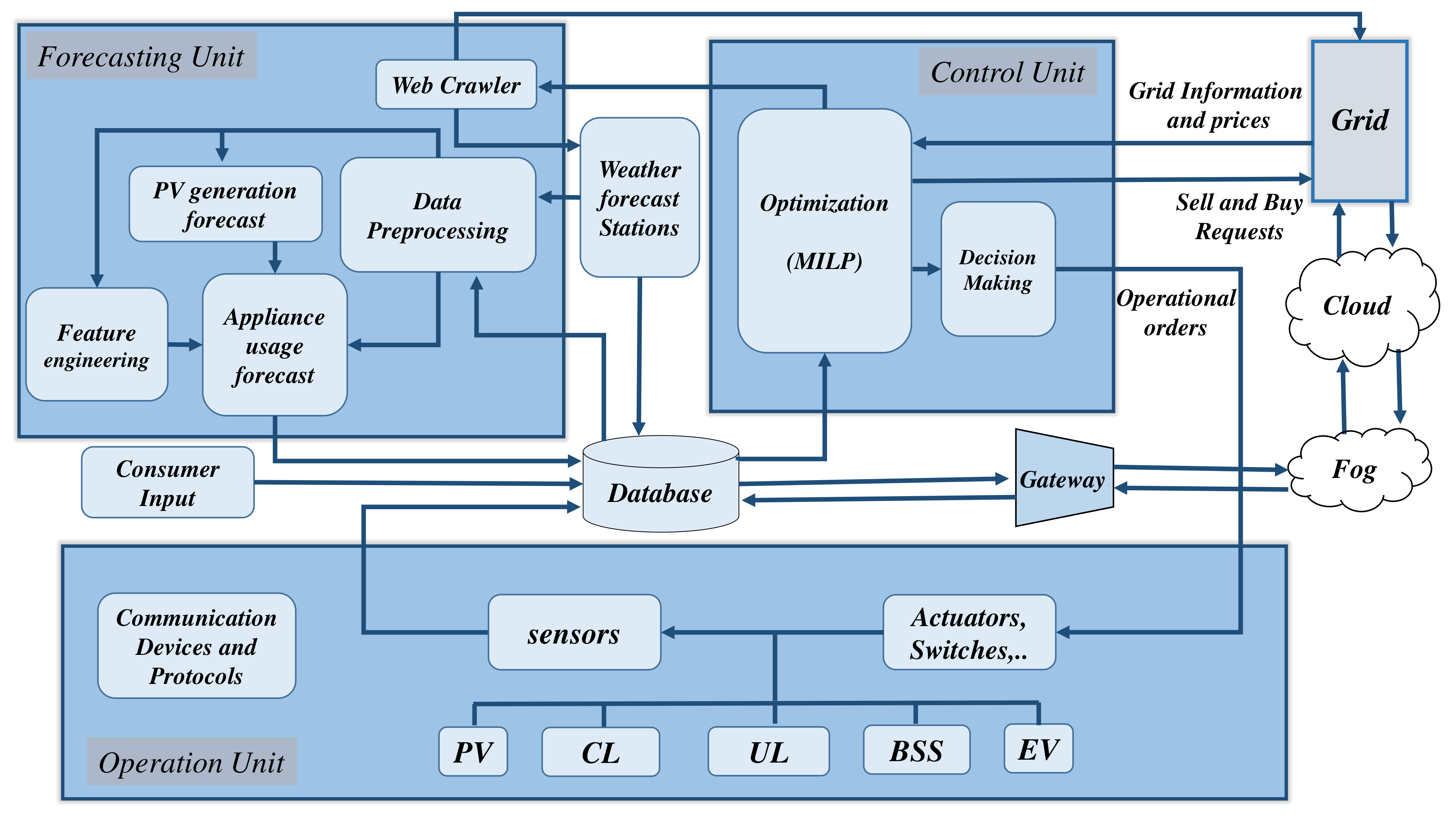}
    \caption{The structure of the proposed management system}
    \label{fig:system_structure}
\end{figure*}

The studied home environment is equipped with controllable and non-controllable electric loads, solar cell production unit, energy storage system, and electric car, all of which are connected to the house's main bus. It is represented according to the three-level architectural model given in the latest version for the reference structure provided by AT\&T, Cisco, General Electric, IBM, and Intel in \cite{26}. An overview of the proposed management system can be seen in Figure \ref{fig:system_structure}.

% \begin{figure}[H]
%     \centering
%     \includegraphics[width=\linewidth]{high_quality_figures/figure_1.png}
%     \caption{The structure of the proposed management system}
%     \label{fig:system_structure}
% \end{figure}

\subsection{Forecasting Unit}
\subsubsection{Data Collection and Pre-processing}
To perform short-term forecasting on the consumption of a household, the 1-minute interval data related to the energy consumption of 4 years in a smart residential house in the city of Austin, Texas, at latitude 30.292432 and longitude -97.699662, have been used. Also, in order to predict the weather conditions, the corresponding recorded weather data were collected for this geographical region. To engage less memory, the data were loaded as float16, which allowed the system to reduce the data size from 630 Mb to 76 Mb. In the following, we will examine the steps of data pre-processing.

\paragraph{Handling Missing Data}
At this stage, the missing data are managed, and the data pass through a noise filter so that the data outside the nominal consumption values of the devices and the minor values are, after ensuring the type of abnormality (physical error or computing system error), removed, interpolated and replaced with appropriate values. Then the values are verified to determine whether they correspond to the possible consumption of the house in terms of statistical criteria or the difference in minutely consumption values. Then the data units are standardized.

\paragraph{Data Normalization}
Prediction models are fed by Features and predictor/independent variables. In this study, environmental data (as independent variables) and time characteristics were given to the model as in Table \ref{tab:correlations}. To increase the prediction accuracy, feature engineering has been used. For example, since electrical energy consumption and consumer behavior can depend on time-based characteristics such as specific holidays or weekends and working days of the week, These features are extracted individually and fed to the model for more accurate prediction.

First, the data is divided into training and testing sets. 85\% of the data was allocated to training and 15\% to testing. Due to the different physical nature and units of these data, some have larger, and some have smaller values. This causes the model to be wrongly biased towards larger numbers. To prevent this, it is necessary to normalize the training and test data. In this study, robust normalization is used to place the values in a comparable range. This normalization scales features that are robust to outliers according to equations \ref{eq:normalization}.

% \begin{equation}
% X^{new} = \frac{X^{old} - Md(X)}{IQR(X)}
% \label{eq:normalization}
% \end{equation}

% \begin{equation}
% IQR(X) = Q3(X) - Q1(X)
% \end{equation}

\begin{subequations}
    \begin{align}
        X^{new} &= \frac{X^{old} - Md(X)}{IQR(X)}
        \label{eq:normalization}
    \end{align}
    \begin{align}
        IQR(X) &= Q_{3}(X) - Q_{1}(X)
    \end{align}
\end{subequations}

In which \(Md(X)\), \(Q_{1}\), \(Q_{3}\), and \(IQR\) are the median, the first quartile, the third quartile of the data, and the interquartile range, respectively.

\subsubsection{Feeding Data to the Model}
After creating the training and test datasets, the input of the models is determined. As input, the models receive 60 batches of training and test data and predict one step (in the case of one minute ahead prediction) or 1440 steps (in the case of the next day prediction) in each iteration.

\subsubsection{Configuration of Predictive Models}
\begin{figure}[H]
    \centering
    \includegraphics[width=\linewidth]{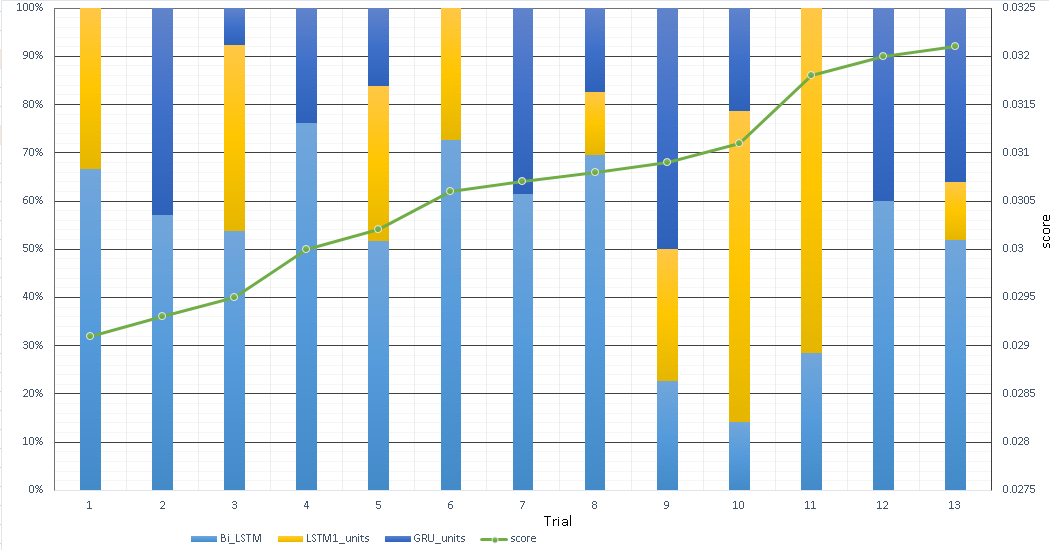}
    \caption{Comparison of superior results from search on hyper-parameters to form optimal deep recurrent model}
    \label{fig:grid_search}
\end{figure}

In order for the predictions to be adaptive to each set of data corresponding to a certain house, First, a grid search is performed on the hyper-parameters of each model when fed with the specific data. This way, the structure for all the practiced models adopts best to each set of collected data based on error criteria. An example of the accumulated grid comparison between the top 13 search results can be seen in the diagram of Figure \ref{fig:grid_search}. After determining the best number of layers and the number of neurons in each layer, the structure of the model is defined and the data is fed to the model. After defining the models, the compilation is done with the "Adam" optimizer based on the mean square error \cite{27}. To compare the models and choose the model with the best and most appropriate performance for our proposed management system, the mean square error criteria has been used. The MSE is great for ensuring that our trained models have no outlier predictions with huge errors, since the MSE puts larger weight on these errors due to the squaring part of the function.
\subsubsection{Prediction Models}
Traditionally, there are various techniques for forecasting using time series data, such as univariate autoregression, univariate moving average, simple exponential smoothing, and especially autoregressive integrated moving average with its various versions and modifications. With the recent progress in computing power of computers, new algorithms have been developed to analyze and forecast time series data. Experimental studies conducted and reported show that algorithms based on deep learning perform better than traditional algorithms such as autoregressive integrated moving \cite{18}.

\begin{table*}[h]
    \centering
    \caption{The correlations between predictors and target variables of prediction}
    \label{tab:correlations}
    \renewcommand{\arraystretch}{1.3} % Increase row spacing for better readability
    \setlength{\tabcolsep}{8pt} % Adjust column spacing

    \begin{tabular}{p{5cm} p{6.5cm} p{2cm}} 
        \hline
        \textbf{Target Variable} & \textbf{Dependent variables and features} & \textbf{$|$correlation$|$} \\
        \hline
        \textbf{Lightings} & Temperature & 0.14 \\
        & Visibility index & 0.31 \\
        & Cloud cover & 0.54 \\
        & Irradiance & 0.80 \\
        & Engineered Time Features & 0.68 \\
        \textbf{Oven} & Engineered Time Features & 0.43 \\
        & Temperature & 0.64 \\
        & Humidity & 0.30 \\
        & Pressure & 0.28 \\
        \textbf{Ice Maker} & Wind Speed & 0.27 \\
        & Precipitation Probability & 0.51 \\
        & Precipitation Intensity & 0.47 \\
        & Engineered Time Features & 0.25 \\
        & Temperature & 0.41 \\
        & Humidity & 0.67 \\
        & Pressure & 0.15 \\
        \textbf{Bathroom} & Wind Speed & 0.11 \\
        & Precipitation Probability & 0.16 \\
        & Precipitation Intensity & 0.38 \\
        & Dew Point & 0.37 \\
        & Engineered Time Features & 0.52 \\
        \textbf{Kitchen (except controllable and lighting loads)} & Engineered Time Features & 0.48 \\
        \textbf{Living room (except thermal and lighting loads)} & Engineered Time Features & 0.54 \\
        \hline
    \end{tabular}
\end{table*}

\textit{Vanishing and exploding gradient.} While the RNN has internal memory to process sequential data, it faces the problem of vanishing gradient when processing long sequences. As the length of the sequence increases during training with gradient descent, the weights of the RNN are updated over time with respect to the weights and the learning rate in the form of back propagation. 

This update shapes from the last hidden layer to the first hidden layer according to the error gradient. Since the absolute value of the partial derivative of a single hidden layer is limited to 1 due to the use of the sigmoid activation function, a long chain of these partial derivatives is multiplied together and causes the gradient to converge to zero according to Equation~(\ref{eq:vanishing_gradient}). 

When the error values are propagated backward from the output layer, the error remains in the unit cell of LSTM networks. The vanishing gradient slows down the training process and heavily biases newer input data, causing the model to forget older loops~\cite{12}.

\begin{equation}
    \lim_{n \to \infty} W^n = 0, \quad W < 1
    \label{eq:vanishing_gradient}
\end{equation}

The early LSTM network, as proposed by Hochreiter and Schmidhuber in 1997~\cite{28}, lacked a forget gate, which is used to remove all unnecessary long-term information. Therefore, there was a vanishing and exploding gradient problem in this network. 

With the introduction of the forget gate, this problem was reduced but not eliminated.

% \begin{figure}[h]
%     \centering
%     \includegraphics[width=0.6\textwidth]{figure2-3.png}
%     \caption{Illustration of the Forget Gate Mechanism in LSTM}
%     \label{fig:forget_gate}
% \end{figure}

In the simple RNN, the gradient of states was calculated as Equation~(\ref{eq:rnn_gradient}), but in LSTM networks, this gradient is calculated as Equation~(\ref{eq:lstm_gradient}). Both sum multipliers have a sigmoid term that can vanish when multiplied $t - t'$ times. However, in a simple RNN, the gradient decays at rate $w \sigma'(\cdot)$, while in LSTM networks, the gradient decays at rate $\sigma(\cdot)$~\cite{19, 20}.

\begin{equation}
    \frac{\partial h_{t'}}{\partial h_t} = \prod_{k=1}^{t - t'} w \sigma' (w h_{t' - k})
    \label{eq:rnn_gradient}
\end{equation}

\begin{equation}
    \frac{\partial CS_{t'}}{\partial CS_t} = \prod_{k=1}^{t - t'} \sigma (v_{t + k})
    \label{eq:lstm_gradient}
\end{equation}
Also, in predicting long sequences using correlated features, not only the error increases, but also all the features should be used to predict two steps in the future in the frame movement to deliver data to the network. The same form is available for the predicted step to be used in the second step. These problems cause us to turn to the encoder- decoder structure (sequence to sequence).

A sequence-to-sequence architecture has been developed for machine learning tasks where both the input and output are sequences. A sequence-to-sequence model consists of three components: an encoder, an intermediate encoding vector, and a decoder. 

Reference~\cite{29} shows that in this model, as the length of the input sequence and the desired sequence for prediction increases, accuracy decreases, and the difference between the output sequence and the desired sequence grows. There is also a bottleneck problem in the sequence-to-sequence structure. 

To improve these conditions, the attention mechanism is used in the sequence-to-sequence model. The attention mechanism was introduced to address the bottleneck problem caused by using a fixed-length encoding vector, where the decoder has limited access to the information provided by the input. This issue is particularly problematic for long or complex sequences, such as our case of study, where their display dimensions must match those of shorter or simpler sequences. 

The attention mechanism is divided into step-by-step calculations of alignment scores, weights, and content vectors. In fact, this mechanism ensures that, through a comparison function, only the parts of the input that are most relevant to our desired prediction are given greater importance, while irrelevant parts of the connection matrix between the inputs, which is a relatively sparse matrix, are ignored. This is done using the attention vector, followed by a feedforward layer.

\subsection{Control Unit}
\subsubsection{Energy Cost Optimization}
\paragraph{Objective Function}
Constrained optimization, as seen in Equation \ref{eq:objective_function}, is done with the objective of minimizing the total daily cost paid to the grid.

\begin{equation}
Minimize \quad  \sum_{t=1}^{T} (P_{t}^{Gr} - P_{t}^{Sell}) \cdot \pi_{t} \cdot \delta_{t}^{step}
\label{eq:objective_function}
\end{equation}

In which \(\pi_{t}\delta_{t}^{step}\), \(P_{t}^{Gr}\), and \(P_{t}^{Sell}\) represent Electric energy price (\$/kWh), the length of each time interval (min), power purchased from the grid at time t (kW) and power sold to the grid at time t (kW), respectively.

\paragraph{Power Balance Constraint}
One of the most important limitations that must be considered in the optimization problem is the power balance equation, which shows that the amount of power produced at any time should be equal to the amount of power consumed at that moment. By dividing the 24 hours into one-minute intervals, we have tried to establish this balance more precisely, thus avoiding the damages caused by minute and sub-minute fluctuations of renewable production power as much as possible. As shown in Equation \ref{eq:power_balance}, when the power required by the smart home is more than the total power produced by solar cells, the power discharged from electric vehicles and energy storage system, this deficiency is inevitably taken from the grid.
\begin{multline}
    P_{t}^{Buy} + P_{t}^{PV,cons} + P_{t}^{EV,cons} + P_{t}^{ESS,cons} = \\
    P_{t}^{EV,ch} + P_{t}^{ESS,ch} + P_{t}^{AC} + P_{t}^{HR} + P_{t}^{NI} \\
    + P_{t}^{Inflex,p} + P_{t}^{Inflex,unp}, \quad \forall t
    \label{eq:power_balance}
\end{multline}

In which \(P_{t}^{Inflex,p}\), \(P_{t}^{Inflex,unp}\), \(P_{t}^{NI}\), \(P_{t}^{HR}\) and \(P_{t}^{AC}\) are power consumed by predictable inflexible loads at time t (kW), power consumed by unpredictable inflexible loads at time t (kW), power consumption of non-interruptible loads at time t (kW), power consumption of the heating load at time t (kW) and power consumption of the cooling load at time t (kW) respectively.

\paragraph{Selling to the Grid}
The presence of renewable production sources in the smart home has made it possible to benefit from selling power to the grid. The power produced by the PV system is stored in the energy storage system, and the excess amount is sold to the grid. Equations \ref{eq:sell_power} - \ref{eq:buy_sell_constraint} express this issue along with the buying and selling restrictions.

\begin{equation}
P_{t}^{sell} = P_{t}^{PV,Sell} + P_{t}^{Ev,Sell} + P_{t}^{ESS,Sell}
\label{eq:sell_power}
\end{equation}

\begin{equation}
P_{t}^{Buy} \leq P_{t}^{Buy,max} \cdot u_{t}^{Buy}, \forall t
\label{eq:buy_constraint}
\end{equation}

\begin{equation}
P_{t}^{sell} \leq P_{t}^{sell,max} \cdot u_{t}^{sell}, \forall t
\label{eq:sell_constraint}
\end{equation}

\begin{equation}
u_{t}^{Buy} + u_{t}^{Sell} \leq 1, \forall t
\label{eq:buy_sell_constraint}
\end{equation}

In which \(P_{t}^{sell}\), \(P_{t}^{Buy,Max}\), \(P_{t}^{Sell,Max}\), and \(u_{t}^{Buy/Sell}\) are the selling power to the grid at time t (kW), the maximum allowed power purchased from the grid at time t (kW), the maximum power allowed to be sold to the network at time t (kW), and binary control variable for buying or selling power from/to the grid, respectively.

\subsection{Operation Unit}
\subsubsection{Communications}
The IoT is a system that has the ability to transmit data through the network without human-to-human or human-to-machine interaction. Internet of Things devices can communicate with each other through wired Ethernet or wireless networks via Bluetooth, Wi-Fi, Z-Wave, ZigBee, or through the cloud or gateway using HTTP, TLS, CoAP, TCP, UDP, or IP protocols \cite{30}. Based on evaluation studies of various communication technologies, due to low cost, low consumption, and ease of use, ZigBee is suitable for smart home management purposes \cite{31}, \cite{32}. For this paper, Zigbee IP is considered, which optimizes the standard for IPv6-based full wireless mesh networks, offering internet connections to control low-power, low-cost devices.

\subsubsection{Load and Power Generation Modeling and Formulation}
This section presents the models of each load participating in the optimization and the componentization of the load scheduling problem. A smart home load scheduling problem can be formulated as an optimization problem considering multiple time intervals, \(\mathbb{T} = \{1, 2, ..., |\mathbb{T}|\}\). The scheduling is formulated as a mixed linear integer programming problem. Each consumer and producer unit at home has its own mathematical optimization models, which include operational constraints on the parameters related to its physical system \cite{33}.

\subsubsection{Classification of Household Loads}
Different authors have categorized loads in different ways. According to the classification of reference \cite{34}, loads are divided into three categories of time-shifting loads, interruptible loads, and weather-based loads. \cite{35} has classified home appliances into two categories: schedulable and non-schedulable. A schedulable device can be interruptible or non-interruptible. For example, a washing machine can be a schedulable, non-interruptible device, and a pool pump can be a schedulable interruptible device. \cite{36} considers the classification of household appliances as uncontrollable, controllable, and schedulable and on/off appliances. In this paper, we consider the classification of household loads as follows:

\paragraph{Unpredictable (Stochastic) Inflexible Loads}
This category includes non-schedulable loads and cannot be easily integrated into energy management plans. Some examples are the consumption of restrooms and laptops.

\paragraph{Predictable Inflexible Loads}
This category includes loads influenced by residents' long-term habits. The general consumption of the kitchen (except controllable appliances), the consumption related to residents' bathing, the consumption of the living room (except controllable appliances), which are collected over the years by smart metering systems, and the consumption of light bulbs are included in this category. These types of data have an utterly random nature in the short term. However, on a larger time scale, by feeding suitable and complex models with big data, a discernible trend can be observed and predicted with partial consideration of time (minutes in a few minutes, a few minutes in an hour, hours in half day, and so on). Pattern recognition of habits, engineering features related to household habits, and using environmental features influencing consumption (such as cloud cover index or irradiance in the case of lighting consumption) help with a more accurate prediction of these loads.

\paragraph{Flexible Loads}
Loads of this category can be scheduled and can participate in the energy management system. Flexible loads can be divided into three categories according to their unique operational characteristics.

\textbf{Interruptible Loads} \hspace{1em}
% \subparagraph{\textbf{Interruptible Loads}}
These loads allow interruption and later resumption without compromising the comfort of the consumer during their operation. Interruptible loads are modeled to consume energy from a specific range, allowing for adjustable power allocation during low and high price periods. Electric vehicles and battery storage systems are considered interruptible loads. These energy storage devices act as power providers during discharge and as consumers during the charging time.

\textbf{Energy Storage System}.
In power management programs, energy storage systems should be modeled according to battery capacity, charging and discharging rates, and efficiency. The equations related to their modeling are presented. Equation \ref{eq:ess_cons} shows that the stored energy can be consumed at home or sold to the grid subject to sales limits. Equations \ref{eq:ess_deh}-\ref{eq:ess_final} show the model and the limits used for energy storage system.

\begin{equation}
-P_{t}^{ESS,deh} \cdot \eta^{ESS} = P_{t}^{ESS,cons} + P_{t}^{ESS,sell}
\label{eq:ess_cons}
\end{equation}

\begin{equation}
-P_{t}^{ESS,ch,max} \cdot (1 - u_{t}^{ESS}) \leq P_{t}^{ESS,deh} \leq 0, \forall t
\label{eq:ess_deh}
\end{equation}

\begin{equation}
R_{t}^{ESS} = R^{ESS,in}, t = 1
\label{eq:ess_initial}
\end{equation}

\begin{equation}
R_{t}^{ESS} = R_{t-1}^{ESS} + (P_{t}^{ESS,ch} \cdot \eta^{EV} + \frac{P_{t}^{ESS,deh}}{\eta^{ESS}}) \cdot \delta t, \forall t > 1
\label{eq:ess_update}
\end{equation}

\begin{equation}
R^{ESS,min} \leq R_{t}^{ESS} \leq R^{ESS,max}, \forall t
\label{eq:ess_capacity}
\end{equation}

\begin{equation}
R_{t}^{ESS} = R_{t}^{ESS,in}, t = T
\label{eq:ess_final}
\end{equation}

\textbf{Electric Vehicle (EV)}. 
Studies have shown that, if properly managed, electric vehicles can be used as flexible loads to provide ancillary services. Managed charging, whether through utilities, load balancers, or aggregators, allows EVs to be used as storage for excess renewable energy generation and to smooth out adverse effects on the net load \cite{37}.

All equations related to electric vehicle modeling are presented in this section \cite{38}. The electric car and the energy storage system have very similar characteristics, with the difference that the electric car is considered disposable. For convenience, it is assumed that the car gets fully charged during the two periods in the parking lot, leaving the house with a full charge. The initial and final capacity of the car battery is different. The car loses some of the stored energy before returning home. The EV model is shown in Equations \ref{eq:ev_ch}-\ref{eq:ev_final}.

% \begin{equation}
% 0 \leq P_{t}^{EV,ch} \leq P_{t}^{EV,ch,max} \cdot u_{t}^{EV}, \forall t \in \{[0, T1] \cup [T2, T]\}
% \label{eq:ev_ch}
% \end{equation}

% \begin{equation}
% -P_{t}^{ESS,deh,max} \cdot (1 - u_{t}^{EV}) \leq P_{t}^{EV,deh} \leq 0, \forall t \in \{(0, T1] \cup [T2, T]\}
% \label{eq:ev_deh}
% \end{equation}

\begin{equation}
\begin{split}
    0 \leq P_{t}^{EV,ch} \leq P_{t}^{EV,ch,max} \cdot u_{t}^{EV}, \\
    \forall t \in \{[0, T1] \cup [T2, T]\}
\end{split}
\label{eq:ev_ch}
\end{equation}

\begin{equation}
\begin{split}
    -P_{t}^\text{ESS,deh,max} \cdot (1 - u_{t}^{EV}) \leq P_{t}^{EV,deh} \leq 0, \\
    \forall t \in \{(0, T1] \cup [T2, T]\}
\end{split}
\label{eq:ev_deh}
\end{equation}

\begin{equation}
R_{t}^{EV} = R^{EV,b}, t = T1
\label{eq:ev_initial}
\end{equation}

% \begin{equation}
% R_{t}^{EV} = R_{t-1}^{EV} + (P_{t}^{EV,ch} \cdot \eta^{EV} + \frac{P_{t}^{EV,deh}}{\eta^{EV}}) \cdot \delta t, \forall t \in \{(0, T1) \cup [T2, T]\}
% \label{eq:ev_update}
% \end{equation}

% \begin{equation}
% R^{EV,min} \leq R_{t}^{EV} \leq R^{EV,max}, \forall t \in \{(0, T1] \cup [T2, T]\}
% \label{eq:ev_capacity}
% \end{equation}

\begin{equation}
\begin{split}
    R_{t}^{EV} = R_{t-1}^{EV} + 
    \Big( P_{t}^{EV,ch} \cdot \eta^{EV} 
    + \frac{P_{t}^{EV,deh}}{\eta^{EV}} \Big) 
    \cdot \delta t, \\
    \forall t \in \{(0, T1) \cup [T2, T]\}
\end{split}
\label{eq:ev_update}
\end{equation}

\begin{equation}
\begin{split}
    R^{EV,min} \leq R_{t}^{EV} \leq R^{EV,max}, \\
    \forall t \in \{(0, T1] \cup [T2, T]\}
\end{split}
\label{eq:ev_capacity}
\end{equation}

\begin{equation}
R_{t}^{EV} = R_{t}^{EV,max}, t = T2
\label{eq:ev_final}
\end{equation}
\textbf{Non-interruptible Loads} \hspace{1em}
% \subparagraph{\textbf{Uninterruptible Loads}}
The operation of these necessary loads differs from interruptible loads. After starting, they must complete a set of predetermined sequential phases. This sequential operation must follow its predefined schedule without interruption. These loads can be modeled to consume a prearranged amount of power during a specified number of consecutive time steps. Washing machines and clothes dryers are examples of this category. These loads cannot be interrupted or changed in the amount of consumption, but the time to turn them on or off can be optimally determined according to the daily electricity pricing \cite{39}. The modeling of these loads has been done by binary coding. For example, the operation of each of the different operating phases of the washing machine is controlled by two binary variables. According to Table \ref{tab:binary_encoding}, equations \ref{eq:binary_constraint1} - \ref{eq:binary_constraint6} are obtained.
\begin{table}[H]
    \centering
    \caption{Truth tables for binary encoding of uninterruptible loads conditions}
    \label{tab:binary_encoding}
    \renewcommand{\arraystretch}{1.2} % Adjusts row height for better readability
    \setlength{\tabcolsep}{5pt} % Adjusts column spacing to fit within margins

    \resizebox{\linewidth}{!}{ % Ensures table fits within the column
    \begin{tabular}{c|c|c|c|c|c|c|c}
        \hline
        $U_{2}^{i,t-1}$ & $U_{2}^{i,t}$ & \textbf{Ok} & $U_{2}^{i,t}$ & $U_{1}^{i,t}$ & $U_{1}^{i,t-1}$ & \textbf{Ok} & \textbf{Not Ok} \\  
        \hline
        0 & 0 & 0 & 0 & 0 & 0 & 0 & 0 \\
        0 & 1 & 1 & 0 & 0 & 0 & 0 & 1 \\
        1 & 0 & $\times$ & 0 & 1 & 0 & 0 & 0 \\
        1 & 1 & 1 & 1 & 0 & $\times$ & 0 & 1 \\
        \hline
        0 & 0 & 1 & 0 & 0 & 1 & 0 & 1 \\
        1 & 0 & 1 & 0 & 1 & 1 & 0 & 1 \\
        1 & 1 & 1 & 1 & 1 & 1 & 1 & 1 \\
        \hline
    \end{tabular}
    }
\end{table}

\begin{equation}
u_{1}^{i,t} + u_{2}^{i,t} \leq 1, \forall i, t
\label{eq:binary_constraint1}
\end{equation}

\begin{equation}
u_{1}^{i,t-1} + u_{1}^{i,t} - u_{2}^{i,t} \leq 0, \forall i, t: t > 1
\label{eq:binary_constraint2}
\end{equation}

\begin{equation}
u_{2}^{i,t-1} - u_{2}^{i,t} \leq 0, \forall i, t: t > 1
\label{eq:binary_constraint3}
\end{equation}

\begin{equation}
u_{2}^{i,t} \leq u_{1}^{i-1,t}, \forall i, t: j > 1, i: i > 1, i < F
\label{eq:binary_constraint4}
\end{equation}

\begin{equation}
\sum_{t=1}^{cl} u_{1}^{i,t} = T^{i}, \forall i, t
\label{eq:binary_constraint5}
\end{equation}

\begin{equation}
\sum_{f=1}^{F} (u_{1}^{i,t} \cdot P_{t}) = P_{t}^{tot}, \forall i, t
\label{eq:binary_constraint6}
\end{equation}

% $U_{1}^{i,t}$ and $U_{2}^{i,t}$ indicate the implementation and the completion of the $i^{\text{th}}$ phase at time $t$, respectively. 

According to the truth Table~\ref{tab:binary_encoding}, the conditions of non-simultaneity of phases and sequence of phases are applied as constraints \ref{eq:binary_constraint1} and \ref{eq:binary_constraint2}, respectively.

In these equations, \(i\), \(U_{1}^{i,t}\), \(U_{2}^{i,t}\), \(T^{i}\), \(P^{i}\), \(P_{t\,tot}^{t}\), and \(F\) represent the set of phases of each device, the binary variable determining the implementation of phase i, the binary variable determining the completion of phase i, the number of time intervals that each phase i lasts, power consumption of each phase i (kW), Power consumption of each device at time t (kW) and the number of phases of each device, respectively.

Also, the need to begin the first phase of the dryer immediately after the last phase of washing is applied with equation \ref{eq:dryer_constraint}.

\begin{equation}
u_{2}^{i_{1},t,WM} - u_{2}^{i_{2},t,Dryer} \leq 0, i_{1} = F^{WM}, i_{2} = 1
\label{eq:dryer_constraint}
\end{equation}

\textbf{Controllable Loads} \hspace{1em}
Loads in this category do not have predetermined sequential phase operation, or their operation cannot be easily resumed later as in the first category because it may violate the desired comfort zone. 

The HVAC system that adjusts the temperature of the indoor environment through control according to the user's comfort temperature range is an example of this category. The heating and cooling systems are modeled as a linear equation that shows the temperature changes in the environment with respect to the consumed power. 

Using the weather forecasts from the outside of the house, the calculation of the internal temperature of the smart house is expressed by a linear relationship that depends on the heat exchange between the house and the outside environment, the thermal parameters of the air, and the thermodynamic parameters of the house. 

The on/off state of the cooler and heater is determined by comparing this calculated temperature and the permitted temperature range. The modeling related to the air conditioning system is presented by equations.

Equations 30a to 30h~\cite{40}:

\begin{subequations}
\begin{align}
    T_{t}^{\text{indoor}} &= T_{t-1}^{\text{indoor}} 
    \left( 1 - \frac{\delta t^{\text{step}}}{1000 \cdot m_{air} \cdot C_{th} \cdot R_{th,eq}} \right) \notag \\
    &\quad + T_{t-1}^{\text{outdoor}} 
    \left(\frac{\delta t^{\text{step}}}{1000 \cdot m_{air} \cdot C_{th} \cdot R_{th,eq}} \right) 
    \label{eq:30a} 
\end{align}
\begin{align}
    M^{AC} &= \delta t^{\text{step}} 
    \left(\frac{EER}{2.77 \times 10^{-4} \cdot m_{air} \cdot C_{th}}\right)
    \label{eq:30b} 
\end{align}
\begin{align}
    M^{HR} &= \delta t^{\text{step}} 
    \left(\frac{COP}{2.77 \times 10^{-4} \cdot m_{air} \cdot C_{th}}\right)
    \label{eq:30c} 
\end{align}
\begin{align}
    \sum_{i=1}^{3} u_{t}^{AC,i} + \sum_{j=1}^{3} u_{t}^{AC,j} &\leq 1, \quad \forall t
    \label{eq:30d} 
\end{align}
\begin{align}
    P_{t}^{AC} &= \sum_{i=1}^{m} P_{AC,j}^{AC,i} u_{t-1}^{AC,i}, \quad \forall t
    \label{eq:30e} 
\end{align}
\begin{align}
    P_{t}^{HR} &= \sum_{j=1}^{n} P_{HR,j}^{HR,j} u_{t-1}^{HR,j}, \quad \forall t
    \label{eq:30f} 
\end{align}
\begin{align}
    \sum_{i=1}^{3} u_{t}^{AC,i} + \sum_{j=1}^{3} u_{t}^{AC,j} &\leq 1, \quad \forall t
    \label{eq:30g} 
\end{align}
\begin{align}
    T_{t}^{\text{indoor,min}} &\leq T_{t}^{\text{indoor}} \leq T_{t}^{\text{indoor,max}}, \quad \forall t
    \label{eq:30h}
\end{align}
\end{subequations}

In which $m_{air}$, $C_{th}$, $R^{th,eq}$, $EER$, and $COP$ are air mass (kg), Heat capacity (kJ/kg$^\circ$F), equivalent of thermal resistance ($^\circ$F/J), energy efficiency ratio, and coefficient of performance, respectively. 

Additionally, they include binary variables $u_{t}^{AC,i}$ and $u_{t}^{HR,j}$ controlling the step of the cooler and heater, step power consumption of the cooler $P_{t}^{AC}$ (kW), and step power consumption of the heater $P_{t}^{HR}$ (kW). 

Furthermore, $T_{t}^{\text{indoor}}$ and $T_{t}^{\text{outdoor}}$ represent the indoor and outdoor temperatures at time $t$, respectively. The minimum and maximum permissible indoor temperatures are denoted as $T_{t}^{\text{indoor,min}}$ and $T_{t}^{\text{indoor,max}}$. The term $\delta t^{\text{step}}$ represents the time step duration, while $M^{AC}$ and $M^{HR}$ are scaling factors for air conditioning and heating, respectively. The variables $P_{AC,j}^{AC,i}$ and $P_{HR,j}^{HR,j}$ denote power-related coefficients in the heating and cooling system calculations.

\subsubsection{Sources of Power Production}
The studied residential house has a PV system as a renewable energy production unit.

The power produced by the solar cell at any moment is obtained using equation \ref{eq:pv_power} \cite{41}. According to equation \ref{eq:pv_balance}, the smart home management system can send a request to buy power or sell it to the grid.

\begin{equation}
P_{t}^{PV} = \eta \cdot S \cdot G_{t}, \forall t
\label{eq:pv_power}
\end{equation}

\begin{equation}
P_{t}^{PV} = P_{t}^{PV,cons} + P_{t}^{PV,sell}, \forall t
\label{eq:pv_balance}
\end{equation}

\section{Scenarios Used to Validate the Proposed Energy Management System}
\label{sec:3}
To validate the model's performance, we consider two scenarios and analyze the scenarios by comparing the cost and fluctuation criteria of the net consumption graph (standard deviation and peak-to-average ratio). The actual consumption of the house based on the forecasts, without applying the proposed energy management system, is considered the initial state. In the first scenario, the operation of the raw management system with shiftable loads, battery and electric vehicle charge control, and thermal control with predetermined non-controllable loads is examined. In the second scenario (Predictive Approach), the performance of the management system with the forecast information of predictable non-controllable loads is tested. Each scenario has been verified using time-of-use and real-time pricing, separately. Table \ref{tab:scenarios} provides an overview of the scenarios.

\begin{table}[H]
    \centering
    \caption{Non-controllable loads in scenarios}
    \label{tab:scenarios}
    \begin{tabular}{l|l|l}
        \textbf{Uncontrollable Loads} & \textbf{Base \& Scenario II} & \textbf{Scenario I} \\
        \hline
        Fig 4 & PA & DA \\
        Fig 5 & Max-Avg & DA \\
        Fig 6 & Median-Avg & \\
    \end{tabular}
\end{table}

Time-of-use pricing is a widely used tariff in which usage costs are divided into different time periods for different seasons of the year or hours of the day. Generally, prices are higher during peak hours and lower during off-peak hours so that consumers can respond to these changes through planning.

The implementation of real-time pricing requires real-time communication between companies and customers and an energy management controller to modify the energy consumption pattern, which reduces the overall price \cite{42}.

\begin{figure}[H]
    \centering
    \includegraphics[width=\linewidth]{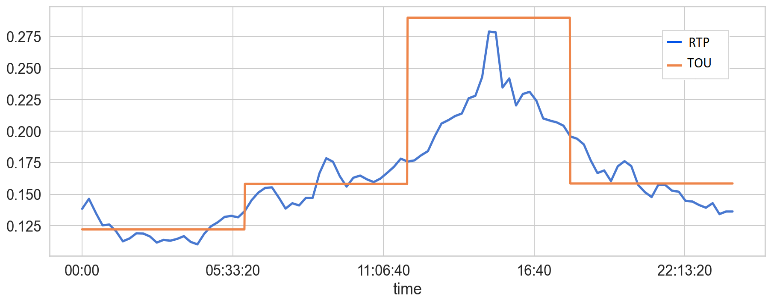}
    \caption{Price data relevant to the studied smart home}
    \label{fig:price_data}
\end{figure}

The relevant price data for the geographic location of the studied smart home was taken online from ERCOT, as presented in Figure \ref{fig:price_data}.

\begin{figure}[H]
    \centering
    \includegraphics[width=\linewidth]{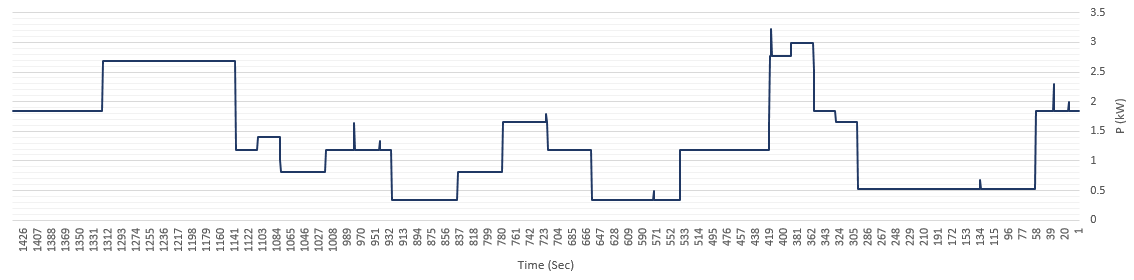}
    \caption{Non-controllable loads based on Max-Mean determination}
    \label{fig:max_avg}
\end{figure}

\begin{figure}[H]
    \centering
    \includegraphics[width=\linewidth]{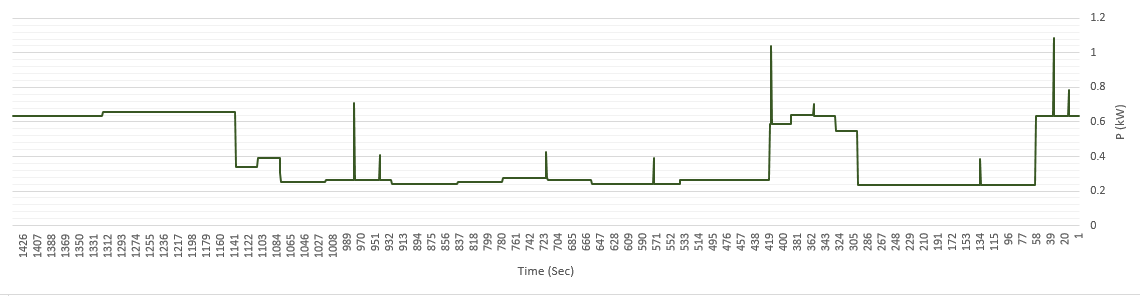}
    \caption{Non-controllable loads based on Median-Mean determination}
    \label{fig:median_avg}
\end{figure}

\begin{figure}[H]
    \centering
    \includegraphics[width=\linewidth]{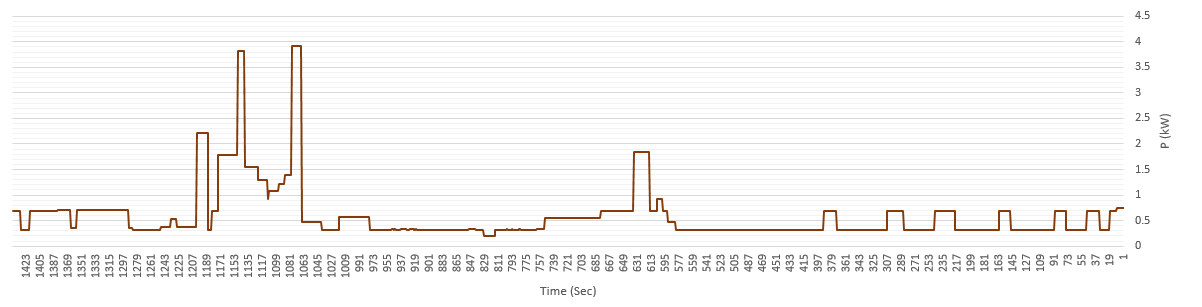}
    \caption{Non-controllable loads based on predictions}
    \label{fig:predicted_loads}
\end{figure}

In the Deterministic Approach (DA) two cases are considered separately. In the first case, the daily average is obtained from the real data of 4 years, and the highest value is considered as the consumption of that load. In the second case, the median of these averages is considered as load consumption. Figures \ref{fig:max_avg}, \ref{fig:median_avg}, and \ref{fig:predicted_loads} show the general diagram of these three loads.

Schedulable loads in the scenarios include clothes washer, dryers, and dishwashers. Controllable loads include battery charging and electric cars. The details of these loads can be seen in Table \ref{tab:schedulable_loads}.

\begin{table}[H]
    \centering
    \caption{Details of Schedulable Loads}
    \label{tab:schedulable_loads}
    \renewcommand{\arraystretch}{1.2} % Adjust row height for readability
    \setlength{\tabcolsep}{6pt} % Adjust column spacing for compactness

    \resizebox{\linewidth}{!}{ % Ensures table fits within the column margins
    \begin{tabular}{l c c}
        \hline
        \textbf{Load Type} & \textbf{Power (kW)} & \textbf{Time (min)} \\
        \hline
        Charging/Discharging Efficiency & 0.95 & - \\
        Maximum EV Capacity & 20 & - \\
        Minimum EV Capacity & 6 & - \\
        Maximum Battery Capacity & 13.5 & - \\
        Maximum Battery Capacity & 4.05 & - \\
        WM, Phase I & 0.15 & 5 \\
        WM, Phase II & 2 & 15 \\
        WM, Phase III & 0.15 & 15 \\
        WM, Phase IV & 2 & 5 \\
        WM, Phase V & 0.15 & 15 \\
        WM, Phase VI & 0.3 & 30 \\
        WM, Phase VII & 0.15 & 5 \\
        Dryer, Phase I & 2.2 & 10 \\
        Dryer, Phase II & 0.15 & 15 \\
        Dryer, Phase III & 2.2 & 10 \\
        DW (Dishwasher), Phase I & 1 & 30 \\
        \hline
    \end{tabular}
    }
\end{table}
  \vspace{-10pt}
The assumptions considered about controllable loads in this study are:
 \vspace{-13pt}
\begin{itemize}
    \item EV connects to the house bus when the residents return home while it has lost half of its charge. The car is a model S Tesla, with level 2 charging and it is fully charged during a period of 7 hours total (according to the capacity and the maximum allowed current of the charging cable).
    \item The battery charge at the first moment of the 24-hour window is equal to the last moment.
    \item The drying operation should start immediately after the washing operation finishes.
    \item There is no delay in the operational phases of the dishwasher, washing machine, or dryer, and the operational phases are carried out continuously one after another.
\end{itemize}

In the optimization of this study, the effective constraint on the thermal load is the range of residents' thermal satisfaction. The thermal load should have the optimal performance in each scenario to have the lowest cost, while keeping the temperature within the allowed range. In this regard, the air conditioning system is installed as a three-step inverter, including three working phases: "Full Load", "Half Load" and "Fan Only". Detailed information about thermal load can be seen in Table \ref{tab:hvac_details}.
\begin{table}[H]
    \centering
    \caption{Details Related to HVAC System}
    \label{tab:hvac_details}
    \renewcommand{\arraystretch}{1.2} % Adjust row height
    \setlength{\tabcolsep}{8pt} % Adjust column spacing

    \begin{tabular}{l c c}
        \hline
        \textbf{Parameter} & \textbf{Unit} & \textbf{Value} \\
        \hline
        Full Load Consumption & kW & 4.2 \\
        Half Load Consumption & kW & 1.9 \\
        Fan Only Consumption & kW & 0.04 \\
        Mass of Air ($M_{air}$) & kg & 1778.369 \\
        Thermal Capacity ($C_{th}$) & kJ/kg°F & 1.01 \\
        Equivalent Thermal Resistance ($R_{th,eq}$) & °F/J & 3.196e-6 \\
        Energy Efficiency Ratio (EER) & - & 3.5 \\
        Coefficient of Performance (COP) & - & 4 \\
        \hline
    \end{tabular}
\end{table}

\section{Results}
\label{sec:4}
\subsection{Comparison of Load Forecasting Results}
After determining the optimal structure for prediction models and feeding data to the model, the results can be seen in Tables \ref{tab:train_test_mse_lighting} and \ref{tab:train_test_mse}.
\begin{table*}[t]
    \centering
    \caption{Train and Test Mean Squared Error (e-4)}
    \label{tab:train_test_mse_lighting}
    \renewcommand{\arraystretch}{1.2} % Increase row height for better readability
    \setlength{\tabcolsep}{6pt} % Adjust column spacing for better fit

    \begin{tabular}{l l | c c | c c | c c | c c | c c }
        \hline
        \textbf{Model} & \textbf{Step} & \multicolumn{2}{c|}{\textbf{External Lighting I}} & \multicolumn{2}{c|}{\textbf{Lighting IV}} & \multicolumn{2}{c|}{\textbf{Lighting III}} & \multicolumn{2}{c|}{\textbf{Lighting II}} & \multicolumn{2}{c}{\textbf{Lighting I}} \\
        &  & Test & Train & Test & Train & Test & Train & Test & Train & Test & Train \\
        \hline
        LSTM   & 1 min   & 2.22  & 2.38  & 0.413  & 0.432  & 7.1   & 0.715  & 0.157  & 1.29   & 8.006  & 3.52   \\
        LSTM   & 1 day   & 3.174 & 2.58  & 0.494  & 0.475  & 7.21  & 0.702  & 0.170  & 1.28   & 8.3    & 5.8    \\
        GRU    & 1 min   & 3.056 & 2.385 & 0.334  & 0.450  & 6.99  & 0.646  & 0.101  & 1.22   & 7.11   & 4.47   \\
        BILSTM & 1 day   & 3.85  & 2.987 & 0.486  & 0.466  & 6.93  & 0.670  & 0.111  & 1.29   & 8.06   & 5.15   \\
        BILSTM & 1 min   & 3.01  & 3.36  & 3.24   & 1.1    & 7.47  & 0.768  & 0.243  & 1.24   & 3.50   & 3.84   \\
        BIGRU  & 1 day   & 3.54  & 3.68  & 0.442  & 0.420  & 7.7   & 0.694  & 0.155  & 1.21   & 6.20   & 4.04   \\
        BIGRU  & 1 min   & -     & -     & 0.101  & 1.28   & 7.18  & 0.856  & 0.186  & 1.39   & -      & -      \\
        SLSTM  & 1 day   & 3.93  & 2.87  & 0.357  & 0.414  & 7.04  & 0.673  & 20.11  & 3.89   & 6.04   & 4.11   \\
        SLSTM  & 1 min   & 1.95  & 1.64  & 6.89   & 5.99   & 4.84  & 0.875  & -      & -      & 11.57  & 5.09   \\
        SGRU   & 1 day   & 2.87  & 1.78  & 9.107  & 5.89   & 9.917 & 1.204  & 10.87  & 2.879  & 18.87  & 5.79   \\
        SGRU   & 1 min   & 1.92  & 1.85  & -      & -      & -     & -      & -      & -      & 8.29   & 3.57   \\
        HRNN   & 1 day   & 3.19  & 2.87  & -      & -      & -     & -      & 0.95   & 0.84   & -      & -      \\
        HRNN   & 1 min   & -     & -     & -      & -      & 18.274 & 1.587 & 0.84   & 0.091  & -      & -      \\
        Seq2Seq & 1 day   & 37.62 & 80.83 & -      & -      & 25.5  & 8.428  & 5.73   & 0.74   & -      & -      \\
        Seq2Seq & 1 min   & 0.048 & 0.085 & 0.053  & 0.0028 & 0.0086 & 0.0084 & -      & -      & 0.00977 & 0.007481 \\
        \hline
    \end{tabular}
\end{table*}

\begin{table*}[t]
    \centering
    \caption{Train and Test Mean Squared Error (e-4)}
    \label{tab:train_test_mse}
    \renewcommand{\arraystretch}{1.2} % Increase row height for better readability
    \setlength{\tabcolsep}{6pt} % Adjust column spacing for better fit

    \begin{tabular}{l l | c c | c c | c c | c c | c c | c c}
        \hline
        \textbf{Model} & \textbf{Step} & \multicolumn{2}{c|}{\textbf{Livingroom}} & \multicolumn{2}{c|}{\textbf{Oven}} & \multicolumn{2}{c|}{\textbf{Ice Maker}} & \multicolumn{2}{c|}{\textbf{Kitchen}} & \multicolumn{2}{c|}{\textbf{Bathroom}} & \multicolumn{2}{c}{\textbf{External Lighting II}} \\
        &  & Test & Train & Test & Train & Test & Train & Test & Train & Test & Train & Test & Train \\
        \hline
        LSTM   & 1 day   & -      & -      & -      & -      & -      & -      & -      & -      & -      & -      & -      & -    \\
        LSTM   & 1 min   & 0.351  & 0.748  & -      & -      & 0.109  & 0.183  & 1.02   & 0.58   & 4.65   & 0.929  & 46.59 & 3.74 \\
        GRU    & 1 day   & 0.231  & 0.734  & -      & -      & 14.84  & 11.76  & 0.957  & 0.842  & 4.75   & 1.09   & 21.79 & 6.77 \\
        GRU    & 1 min   & -      & -      & -      & -      & 0.092  & 0.172  & 1.204  & 0.687  & 5.10   & 1.34   & 20.97 & 5.33 \\
        BILSTM & 1 day   & 2.24   & 0.970  & -      & -      & 14.84  & 11.76  & 1.039  & 0.840  & 4.64   & 1.28   & 23.74 & 7.17 \\
        BILSTM & 1 min   & 0.343  & 0.743  & -      & -      & 0.112  & 0.175  & 1.25   & 1.05   & -      & -      & 46.01 & 3.60 \\
        BIGRU  & 1 day   & 0.163  & 0.712  & -      & -      & 0.102  & 0.182  & 1.388  & 1.28   & 5.51   & 0.751  & 20.11 & 3.89 \\
        BIGRU  & 1 min   & -      & -      & -      & -      & 0.116  & 0.180  & 1.98   & 0.978  & -      & -      & -     & -    \\
        SLSTM  & 1 day   & 5.39   & 1.62   & -      & -      & 14.84  & 11.76  & 2.05   & 0.954  & 4.34   & 1.45   & 22.80 & 4.35 \\
        SLSTM  & 1 min   & -      & -      & -      & -      & 0.85   & 0.47   & 8.169  & 0.59   & 15.62  & 3.85   & 15.62 & 3.85 \\
        SGRU   & 1 day   & -      & -      & -      & -      & 8.49   & 8.45   & 0.87   & 0.74   & -      & -      & 18.59 & 3.94 \\
        SGRU   & 1 min   & 1.85   & 0.829  & -      & -      & -      & -      & -      & -      & 0.85   & 0.015  & -     & -    \\
        HRNN   & 1 day   & -      & -      & -      & -      & 0.68   & 0.49   & 5.87   & 2.845  & -      & -      & -     & -    \\
        HRNN   & 1 min   & -      & -      & 0.284  & 0.0158 & -      & -      & -      & -      & -      & -      & -     & -    \\
        Seq2Seq & 1 day   & -      & -      & 0.102  & 0.0957 & 0.153  & 0.194  & -      & -      & 11.58  & 1.16   & 50.02 & 8.29 \\
        Seq2Seq & 1 min   & -      & -      & -      & -      & 0.119  & 0.0824 & -      & -      & -      & 0.098  & -     & -    \\
        \hline
    \end{tabular}
\end{table*}

\begin{table*}[t]
    \centering
    \caption{Numerical results from the optimizations in different scenarios}
    \label{tab:scenario_results}
    \renewcommand{\arraystretch}{1.3} % Adjust row height for readability
    \setlength{\tabcolsep}{6pt} % Adjust column spacing

    \begin{tabular}{|l|c|c|cc|c|}
        \hline
        \textbf{Category} & \textbf{Tariff} & \textbf{Base} & \multicolumn{2}{c|}{\textbf{Scenario I}} & \textbf{Scenario II} \\
        \hline
        % Row for Uncontrollable Loads
        \multirow{2}{*}{\textbf{Uncontrollable Loads}} 
            &  & PA & \multicolumn{2}{c|}{DA} & PA \\
            &  &  & Max-Avg & Median-Avg & \\
        \hline

        % Controllable Loads (removed upper sub-row)
        \multirow{2}{*}{\textbf{Controllable Loads}}
            & TOU & Fig 13 & Fig 14 & Fig 15 & Fig 18 \\
            & RTP &  & Fig 16 & Fig 17 & Fig 19 \\
        \hline

        % Thermal Load
        \multirow{2}{*}{\textbf{Thermal Load}}
            & TOU & - & Fig 20 & Fig 21 & Fig 24 \\
            & RTP & - & Fig 22 & Fig 23 & Fig 25 \\
        \hline

        % Sell/Buy to/from Grid
        \multirow{2}{*}{\textbf{Sell/Buy to/from Grid}}
            & TOU & - & Fig 26 & Fig 27 & Fig 30 \\
            & RTP & - & Fig 28 & Fig 29 & Fig 31 \\
        \hline

        % PAR of Net Consumption
        \multirow{2}{*}{\textbf{PAR of Net Consumption}}
            &  TOU   & 8.9215  & \textbf{4.9788} & 6.0953 & 7.0570 \\
            & RTP & 
            &6.4854 &   7.3775 &   8.8359 \\
        \hline

        % SD of Net Consumption
        \multirow{2}{*}{\textbf{SD of Net Consumption}}
            & TOU  & 2.9340  & 2.4803 & 2.4006 & \textbf{2.3559} \\
            & RTP &      & 2.7576 & 2.8393 & 2.7986 \\
        \hline

        % Daily Net Cost
        \multirow{2}{*}{\textbf{Daily Net Cost}}
            & TOU & 4.8677 & 4.4651 & 3.5119 &
     \textbf{1.8472}\\
            & RTP & 4.9531 &
            4.5212&
            3.6028 &  2.1155  \\
        \hline
    \end{tabular}
\end{table*}

In these predictions, the important point is to ensure that the models are not over/under-fitted. Overfitting occurs when the model understands the features and characteristics of the training data very well and therefore has a biased performance and does not perform well on the test data. This can be recognized from the learning curve. The learning curve is a graph that records the training error and test error at each epoch of model training. If the curve of training error is much lower than that of test error, the model is overfitted. To solve this problem, in this paper, the random dropout rate in the dropout layer is increased, or in deeper networks, dropout layers are inserted between other layers. Also, underfitting occurs when either the data is not a good representative of the features or the model is not given enough information. In this case, the model will not perform well on the training data. This is recognizable in the learning curve if the training error curve is much higher than the test error curve, and the two graphs do not converge properly. Finally, the learning graph should not have too many dynamic or long-term fluctuations.

\begin{figure}[H]
    \centering
    \includegraphics[width=\linewidth]{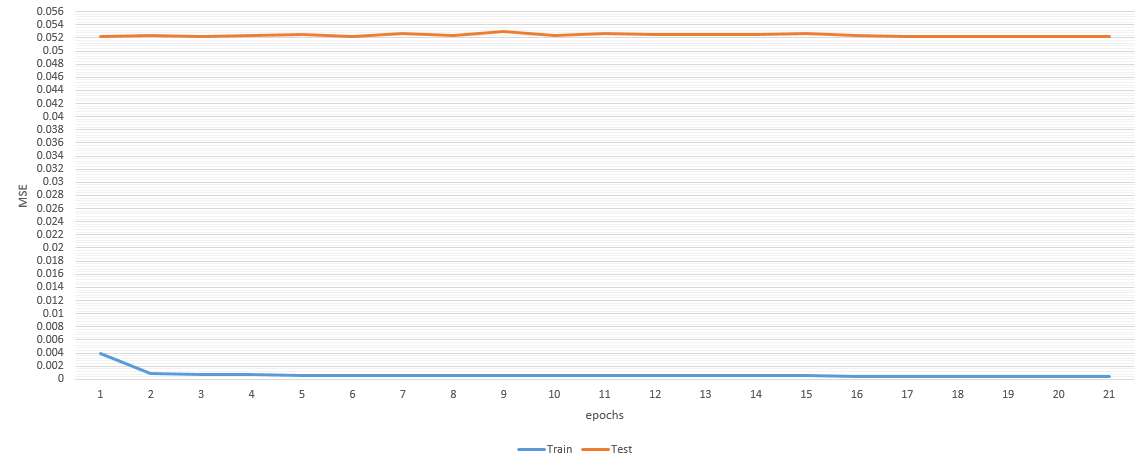}
    \caption{An example of an overfitting in learning graph with early stopping (Bi-Gru model, external lighting II data)}
    \label{fig:overfitting}
\end{figure}

\begin{figure}[H]
    \centering
    \includegraphics[width=\linewidth]{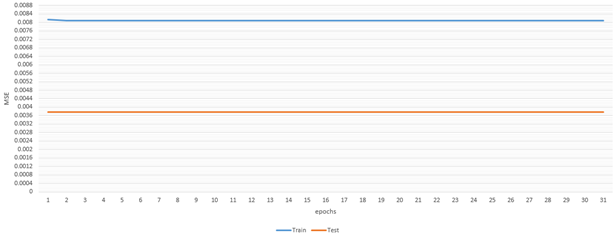}
    \caption{An example of an underfitting in learning graph with early stopping (HRNN model, external lighting I data)}
    \label{fig:underfitting}
\end{figure}

\begin{figure}[H]
    \centering
    \includegraphics[width=\linewidth]{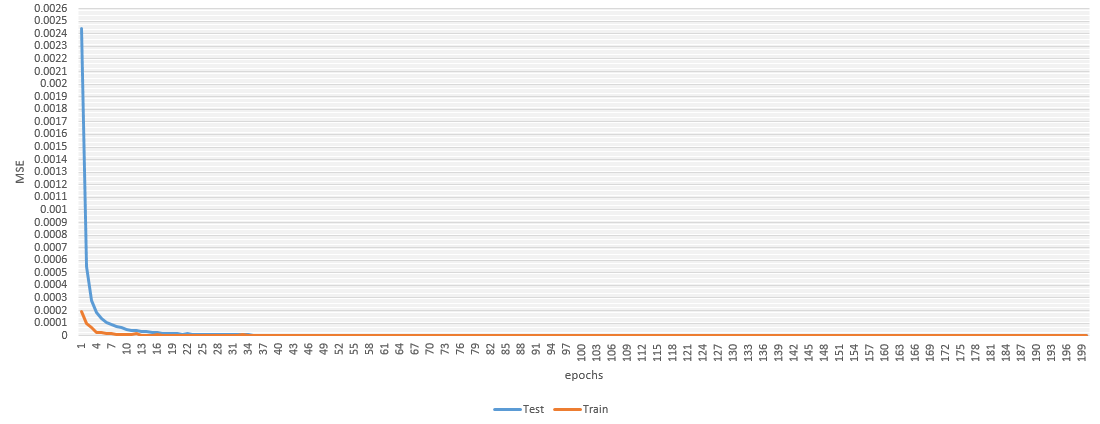}
    \caption{An example of a properly fitted learning graph without early stopping (seq2seq model, lighting II data)}
    \label{fig:proper_fit}
\end{figure}

For all the results mentioned in Tables \ref{tab:train_test_mse_lighting} and \ref{tab:train_test_mse}, the learning graph was drawn, and in cases of underfitting or overfitting, the problems of the model or the data have been fixed. Empty boxes in these tables  are related to cases where these problems were not resolved by any means, indicating that the model was not particularly suitable for the data, so no results were recorded. An example of learning curves suggesting an overfitted or underfitted model can be seen in Figures \ref{fig:overfitting} and \ref{fig:underfitting}. The most appropriate fit is as the sample diagram in Figure \ref{fig:proper_fit}.

All the training processes were executed on a GPU GTX 3090 with 12 cores and 48 GBs of RAM in the JupyterLab environment. Each iteration took 0.1 to 0.5 minutes on average while this number was 80 times bigger when tested on a CPU with 26GBs of RAM. The number of iterations was selected as 200 by default, but us- ing early stopping with a "patience" of 15 and a learning rate reducer factor of 0.5 to reduce the learning rate if the error did not improve within 15 iterations, made this operation efficient and the number of iterations different for each model. The batch size was set to 1000.
As can be seen from Tables \ref{tab:train_test_mse_lighting} and \ref{tab:train_test_mse}, the predic- tion error for both one-minute and one-day pre- dictions, as well as the difference between these two errors, is much lower in the case of the seq2seq model than other models. Therefore, it is much more suitable for day-ahead forecasting.
\subsection{The Results of Cost Optimization and Load Scheduling}
After accurately forecasting predictable uncontrollable loads and calculating the errors, we plan the consumption of the smart home according to these loads to study the effect of these loads on the total cost and consumption fluctuations.

As mentioned before, Adding large amounts of renewable energy to the power grid creates a high level of unpredictability and uncertainty. These challenges are addressed in this work using IoT technology to form a management system.

\begin{figure}[H]
    \centering
    \includegraphics[width=\linewidth]{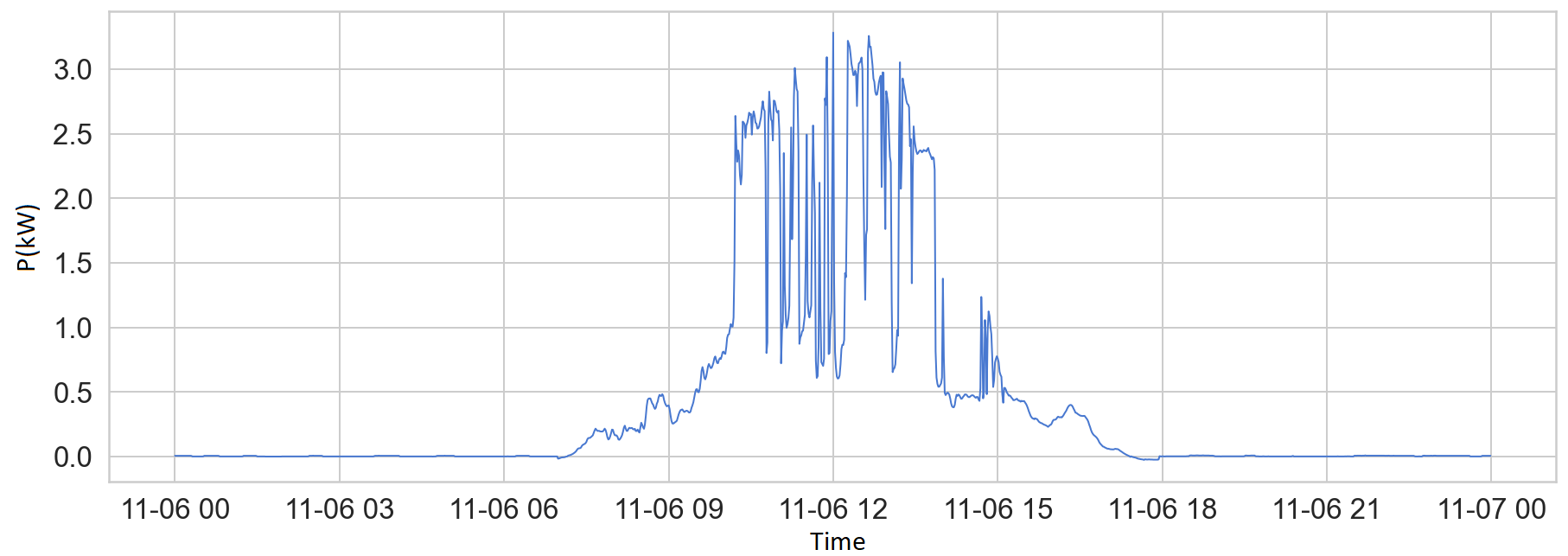}
    \caption{Solar production in 24 hours}
    \label{fig:solar_production}
\end{figure}

\begin{figure}[H]
    \centering
    \includegraphics[width=\linewidth]{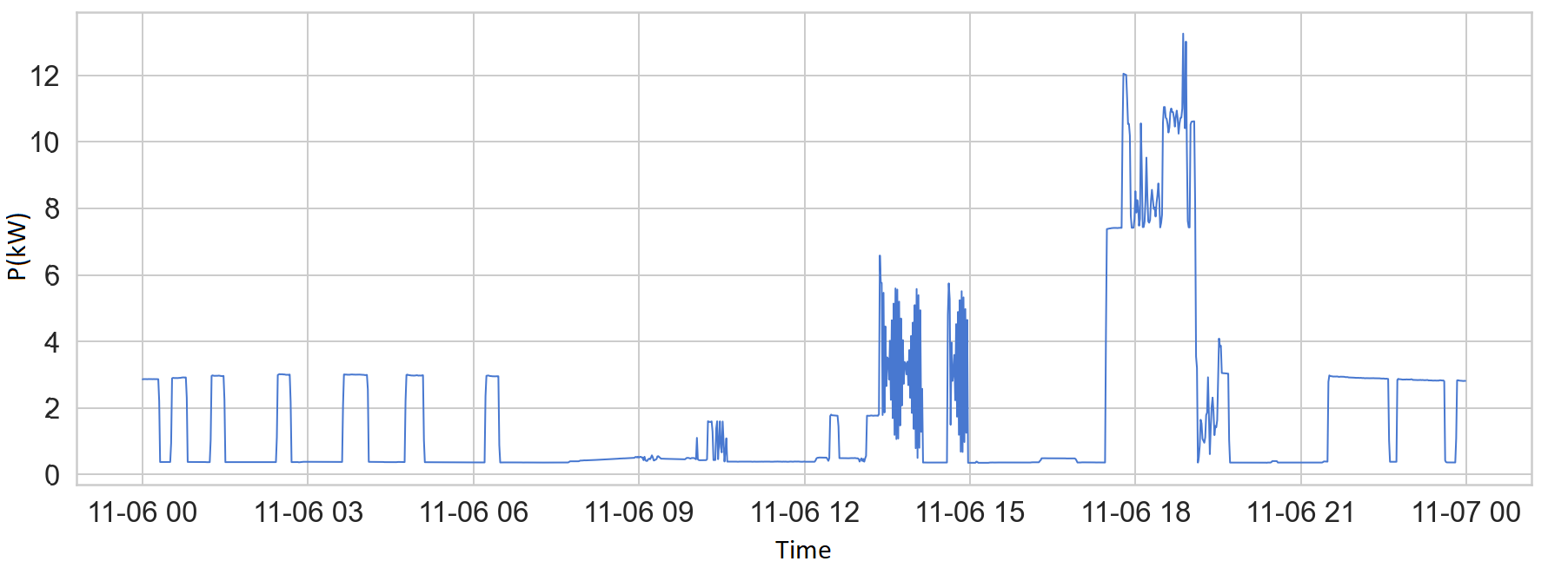}
    \caption{Total consumption of the smart home in 24 hours}
    \label{fig:total_consumption}
\end{figure}

\begin{figure}[H]
    \centering
    \includegraphics[width=\linewidth]{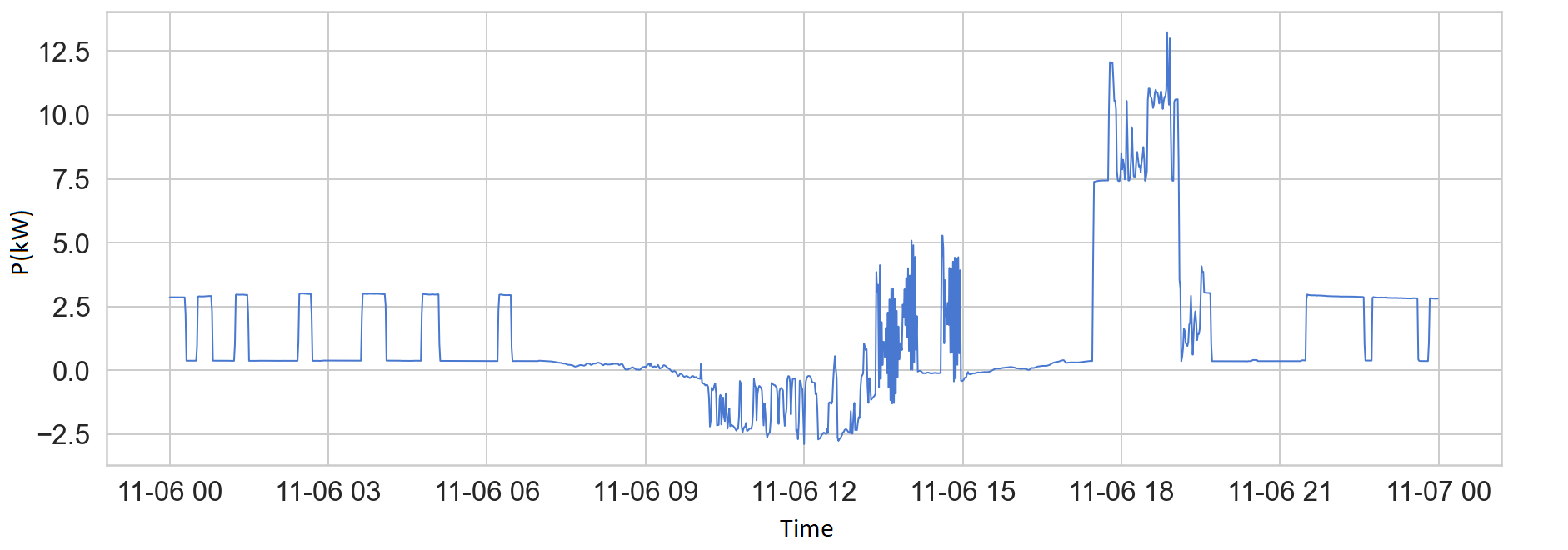}
    \caption{Total net consumption of the smart home in 24 hours}
    \label{fig:net_consumption}
\end{figure}

In the case of this study, according to the raw data, the solar production, demand and net consumption of the house are shown in Figures \ref{fig:solar_production}, \ref{fig:total_consumption}, and \ref{fig:net_consumption}. Negative values indicate the return of current to the grid. In these figures, the time mismatch between consumption and production and the increased fluctuations are noticeable, which have increased the standard deviation (SD) from 2.635 Wh to 2.934 Wh.

SD is an index to measure the deviation of the curve from the average value (equation \ref{eq:sd}) \cite{43}. Adding a solar system to the smart home under study caused a 10.19\% increase in the consumption SD, which means that the house is unsuccessful in matching supply with demand in a way that does not violate system stability. Also, the PAR index represents the ratio of peak electricity consumption to the average value (equation \ref{eq:par}). Limiting the PAR index, like the SD, is important for both the grid and the users because it smooths the load curve for the service providers and stops the operation of the reserve power plants during peak hours, thus reducing the energy cost for the consumers.

\begin{equation}
P_{t}^{Net} = P_{t,N}^{cons} - P_{t}^{solar} - P_{t}^{EV,deh} - P_{t}^{ESS,deh}
\label{eq:net_consumption}
\end{equation}

\begin{equation}
SD = \sqrt{\frac{1}{1440} \sum_{t=1}^{1440} (P_{t}^{Net} - P_{avg}^{Net})^{2}}
\label{eq:sd}
\end{equation}

\begin{equation}
PAR = \frac{\max P_{t}^{Net}}{\frac{1}{1440} \sum_{t=1}^{1440} P_{t}^{Net}}, \quad 0 < t < 1440
\label{eq:par}
\end{equation}

In these equations, $P_{t}^{Net}$, $P_{avg}^{Net}$, and $N$ are the net consumption of the smart home at time $t$ (kW), the average net consumption of the smart home (kW), and the number of smart home electricity consumption sets, respectively.

\subsection{The Results of the Scenarios}
After applying the conditions of the scenario and recording the numerical results related to PAR, SD, and daily net cost, the optimization results in each scenario are presented in Table~\ref{tab:scenario_results}.

\subsection{Results Related to Controllable and Schedulable Loads}
\begin{figure}[H]
    \centering
    \includegraphics[width=\linewidth]{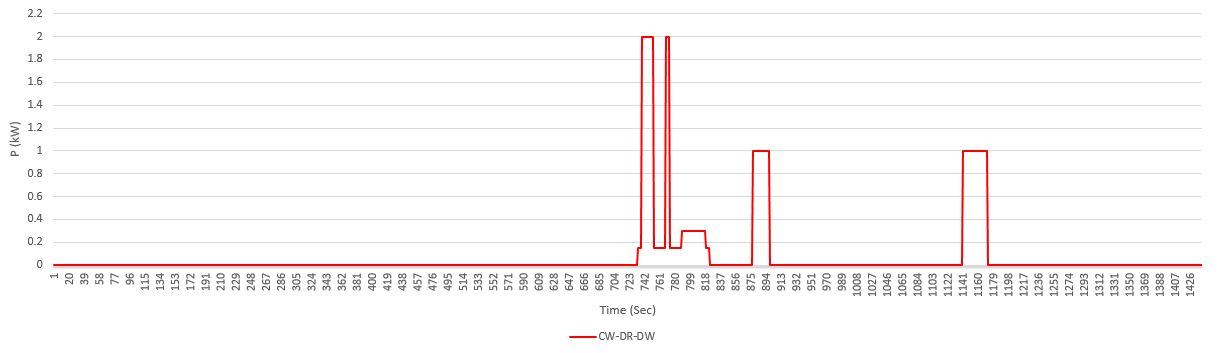}
    \caption{Schedulable loads, scenario I with Max-Avg Deterministic Approach, TOU pricing.}
    \label{fig:scenario1_tou}
\end{figure}

\begin{figure}[H]
    \centering
    \includegraphics[width=\linewidth]{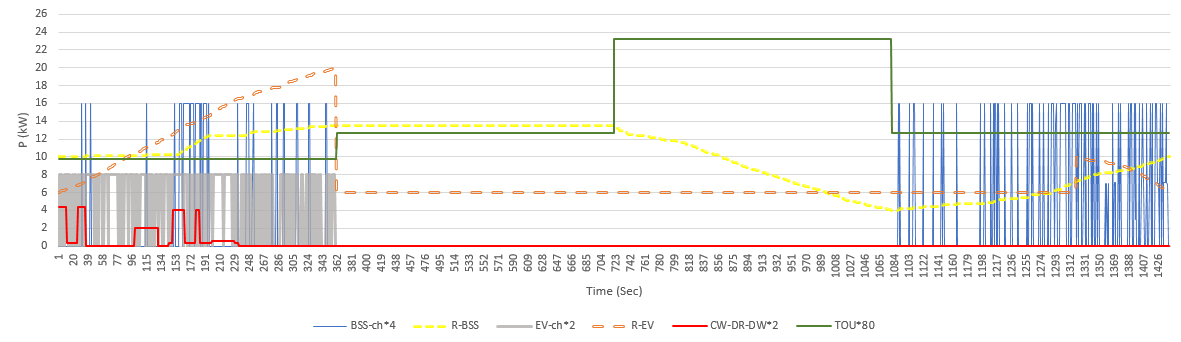}
    \caption{Schedulable loads, scenario I with Max-Median Deterministic Approach, TOU pricing.}
    \label{fig:scenario1_median_tou}
\end{figure}

\begin{figure}[H]
    \centering
    \includegraphics[width=\linewidth]{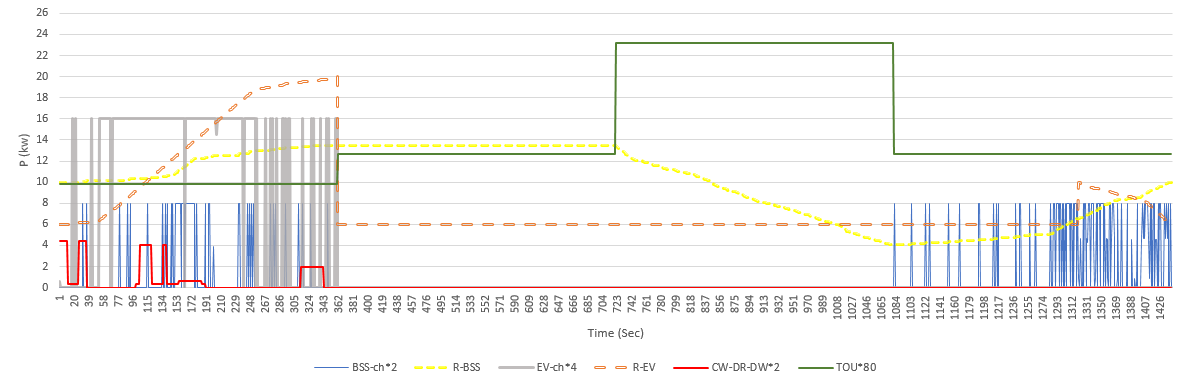}
    \caption{Schedulable loads, scenario I with Max-Avg Deterministic Approach, RTP.}
    \label{fig:scenario1_rtp}
\end{figure}

\begin{figure}[H]
    \centering
    \includegraphics[width=\linewidth]{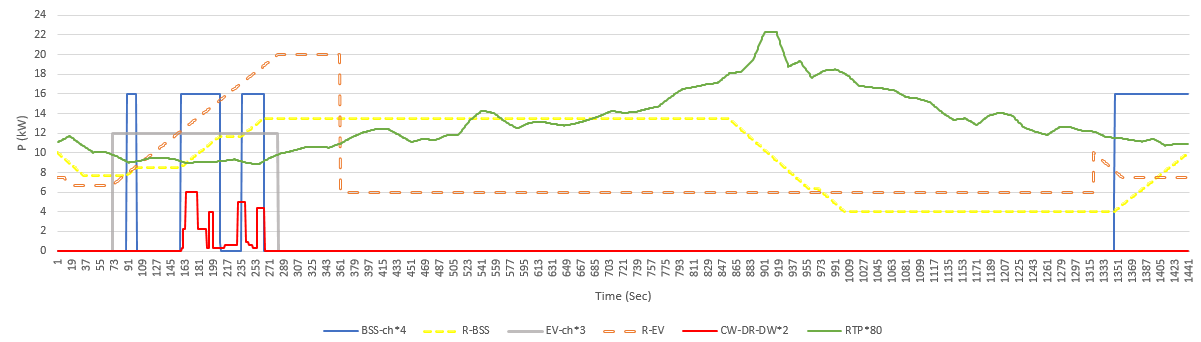}
    \caption{Schedulable loads, scenario I with Max-Median Deterministic Approach, RTP.}
    \label{fig:scenario1_median_rtp}
\end{figure}

\begin{figure}[H]
    \centering
    \includegraphics[width=\linewidth]{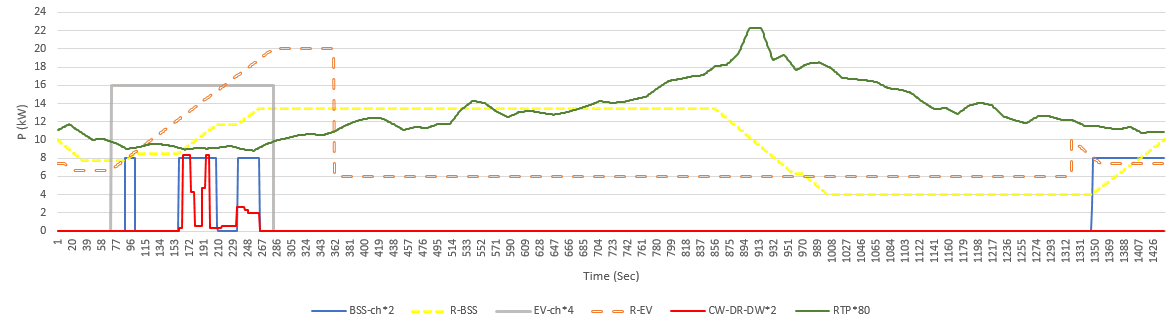}
    \caption{Schedulable loads, scenario II, TOU pricing.}
    \label{fig:scenario2_tou}
\end{figure}

\begin{figure}[H]
    \centering
    \includegraphics[width=\linewidth]{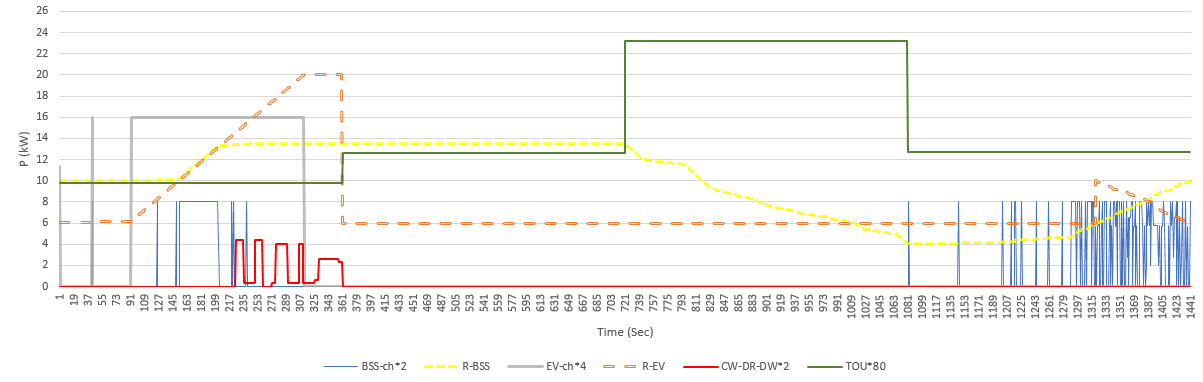}
    \caption{Schedulable loads, scenario II, RTP.}
    \label{fig:scenario2_rtp}
\end{figure}

With the conditions introduced for each scenario, the distribution of schedulable loads during 1440 minutes is obtained, as shown in \ref{fig:scenario1_tou} - \ref{fig:scenario2_rtp}. According to these graphs, the proposed management system in both scenarios has succeeded in optimally transferring the consumption of the washing machine, dryer, and dishwasher with specified limits to the times when the price of electricity is lower. It has also tried to charge the electric car and battery at certain times with lower electricity prices.

% In these equations, \(P_{t}^{Net}\), \(P_{avg}^{Net}\), and \(N\) are the net consumption of the smart home at time \(t\) (kW), the average net consumption of the smart home (kW), and the number of smart home electricity consumption sets, respectively.
% \subsection{The Results of the Scenarios}
% After applying the conditions of the scenario and recording the numerical results related to PAR, SD, and daily net cost, the optimization results in each scenario are presented in Table \ref{tab:scenario_results}.
% \subsection{Results Related to Controllable and Schedulable Loads}
% With the conditions introduced for each scenario, the distribution of schedulable loads during 1440 minutes is obtained, as shown in Figures \ref{fig:scenario1_tou} - \ref{fig:scenario2_rtp}. According to these graphs, the proposed management system in both scenarios has succeeded in optimally transferring the consumption of the washing machine, dryer, and dishwasher with specified limits to the times when the price of electricity is lower, and it has also tried to charge the electric car and battery at certain times with lower electricity prices.

\subsection{Results Related to Thermal Load}
According to the range of thermal comfort determined by users, Figures \ref{fig:thermal_scenario1_median_tou} - \ref{fig:thermal_scenario2_rtp} show that the proposed system has managed to keep the temperature in the allowed range in all scenarios in such a way that, according to the limits and performance steps determined, leads to the lowest daily cost for the consumer (lowest in Fig.~\ref{fig:thermal_scenario2_rtp}).

\begin{figure}[H]
    \centering
    \includegraphics[width=\linewidth]{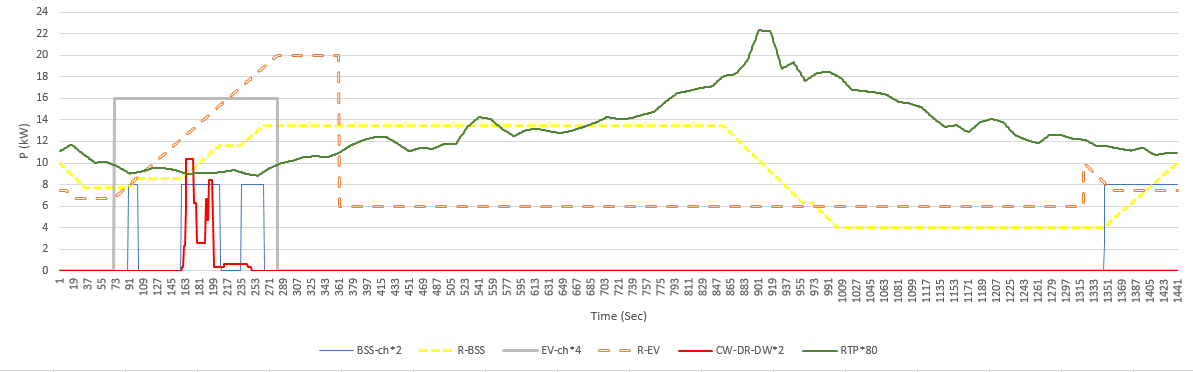}
    \caption{Thermal load, scenario I with Max-Avg Deterministic Approach, TOU pricing.}
    \label{fig:thermal_scenario1_tou}
\end{figure}

\begin{figure}[H]
    \centering
    \includegraphics[width=\linewidth]{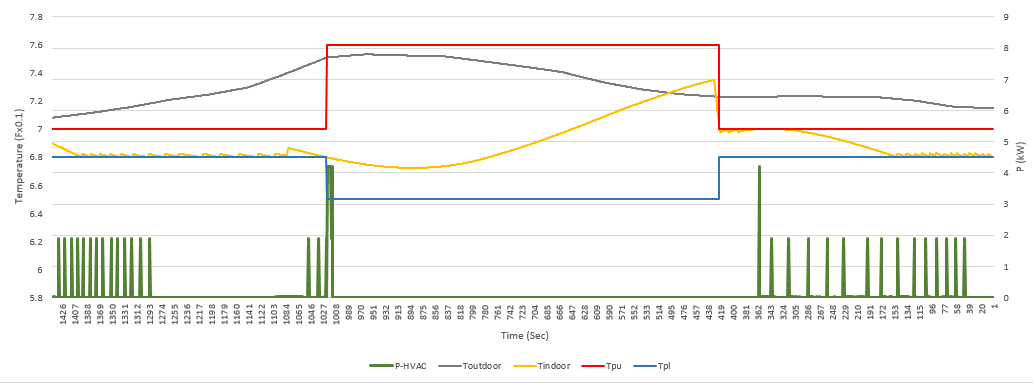}
    \caption{Thermal load, scenario I with Max-Avg Deterministic Approach, TOU pricing.}
    \label{fig:thermal_scenario1_median_tou}
\end{figure}

\begin{figure}[H]
    \centering
    \includegraphics[width=\linewidth]{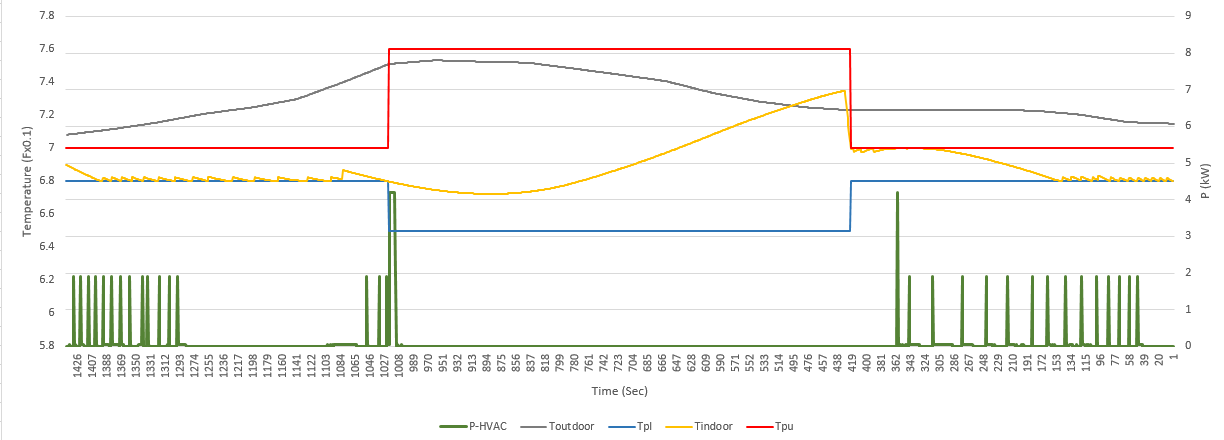}
    \caption{Thermal load, scenario I with Max-Median Deterministic Approach, TOU pricing.}
    \label{fig:thermal_scenario1_rtp}
\end{figure}

\begin{figure}[H]
    \centering
    \includegraphics[width=\linewidth]{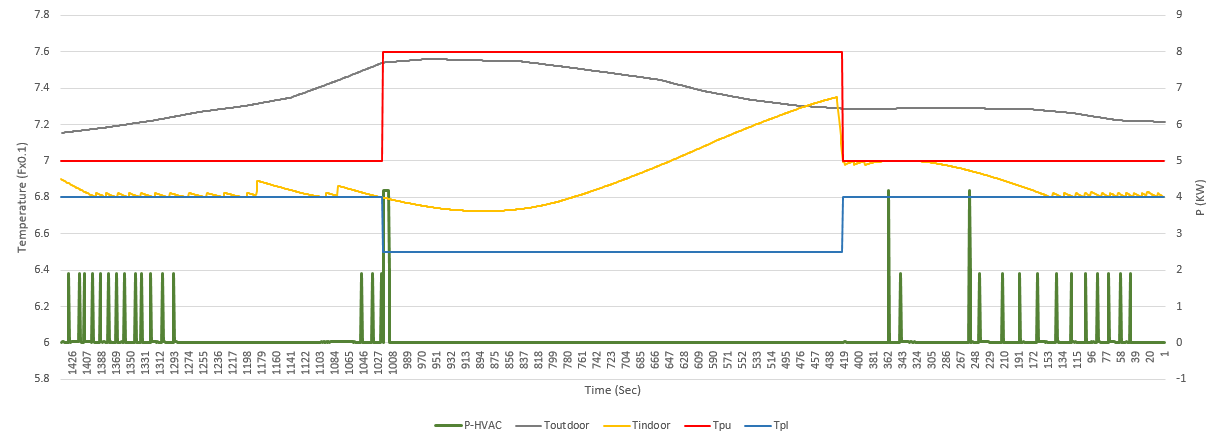}
    \caption{Thermal load, scenario I with Max-Avg Deterministic Approach, RTP.}
    \label{fig:thermal_scenario1_median_rtp}
\end{figure}

\begin{figure}[H]
    \centering
    \includegraphics[width=\linewidth]{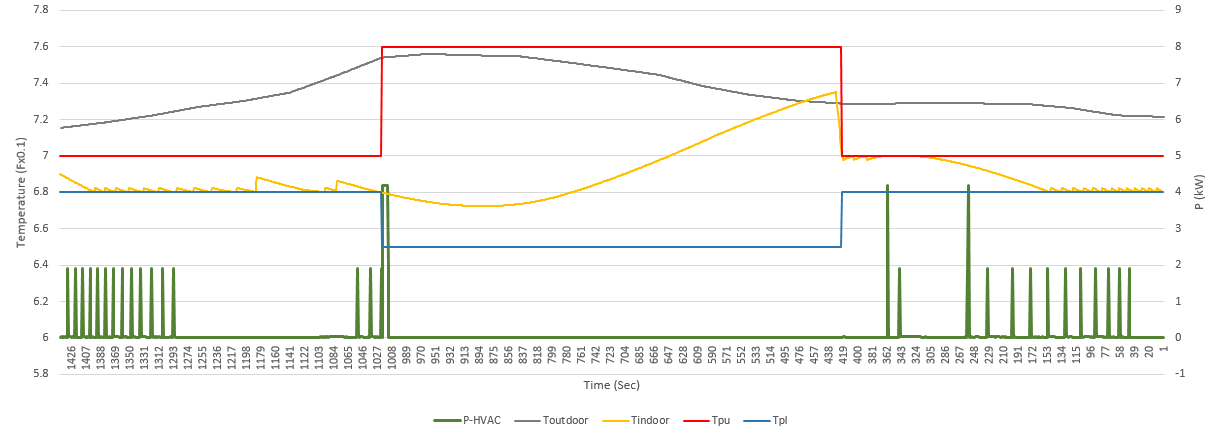}
    \caption{Thermal load, scenario I with Max-Median De- terministic Approach, RTP.}
    \label{fig:thermal_scenario2_tou}
\end{figure}

\begin{figure}[H]
    \centering
    \includegraphics[width=\linewidth]{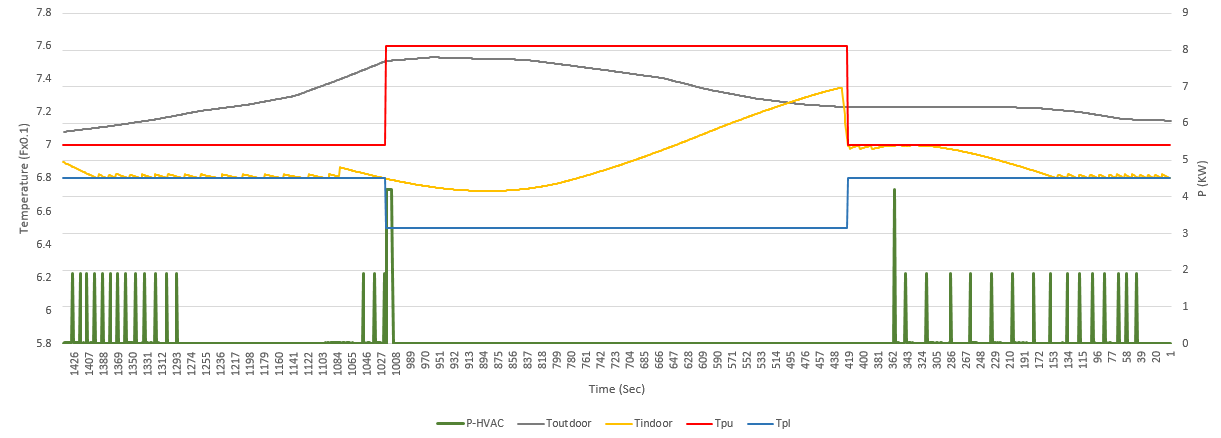}
    \caption{Thermal load, scenario II, TOU pricing.}
    \label{fig:thermal_scenario2_rtp}
\end{figure}

\begin{figure}[H]
    \centering
    \includegraphics[width=\linewidth]{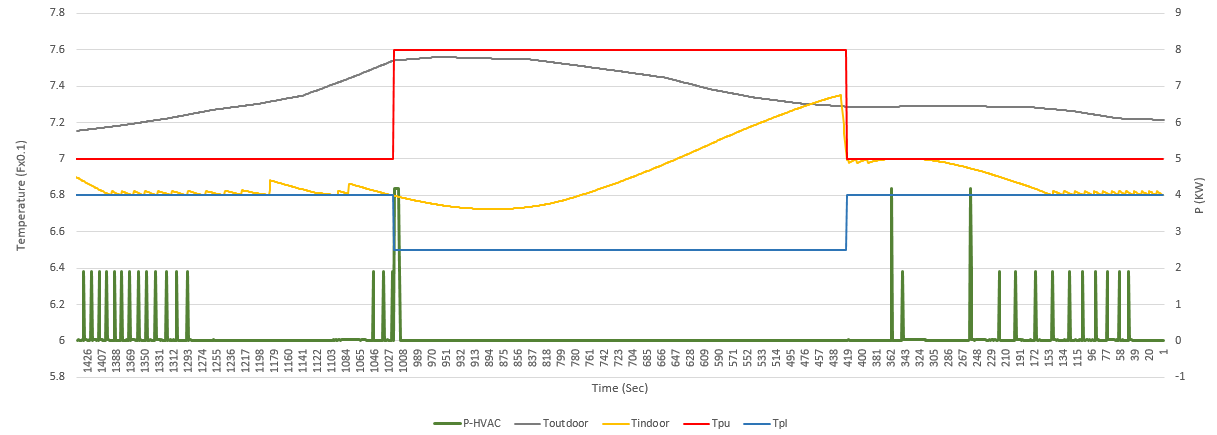}
    \caption{Thermal load, scenario II, RTP.}
    \label{fig:buy_sell_scenario1_tou}
\end{figure}

\subsection{Results Related to Buying and Selling from/to the Grid}
According to Figures \ref{fig:buy_sell_scenario1_tou} - \ref{fig:buy_sell_scenario2_rtp}, the proposed management system has succeeded, during 24 hours, to use internal resources in times of higher electricity price and to establish a balance between internal consumption and purchase from the grid in lower price zones, according to the optimization, in such a way that the amount of total sales to the grid and the charging of the electric car and battery is done more during these times. These results are true for both real-time pricing and time-of-use pricing. As can be seen in Figures \ref{fig:buy_sell_scenario1_rtp} - \ref{fig:buy_sell_scenario2_rtp}, in real-time pricing, in the early hours of the morning, the price of electricity passes a local peak. Therefore, the management system has planned to sell electricity to the grid and use internal resources. It can be seen that by managing internal resources with the proposed system, during times of peak electricity prices, purchasing from the grid is minor and close to zero.

\begin{figure}[H]
    \centering
    \includegraphics[width=\linewidth]{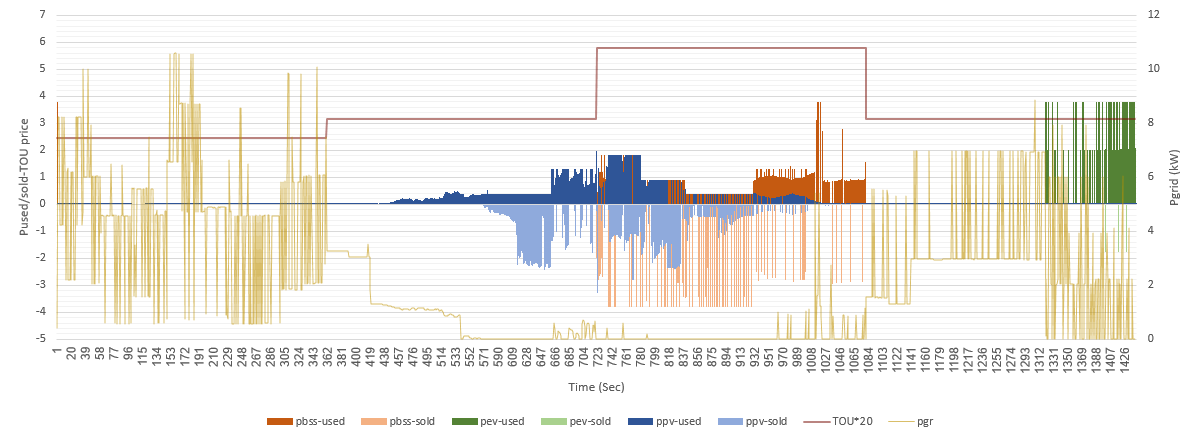}
    \caption{The amount of buying/selling from/to the grid, scenario I with Max-Avg Deterministic Approach, TOU pricing.}
    \label{fig:buy_sell_scenario1_tou}
\end{figure}

\begin{figure}[H]
    \centering
    \includegraphics[width=\linewidth]{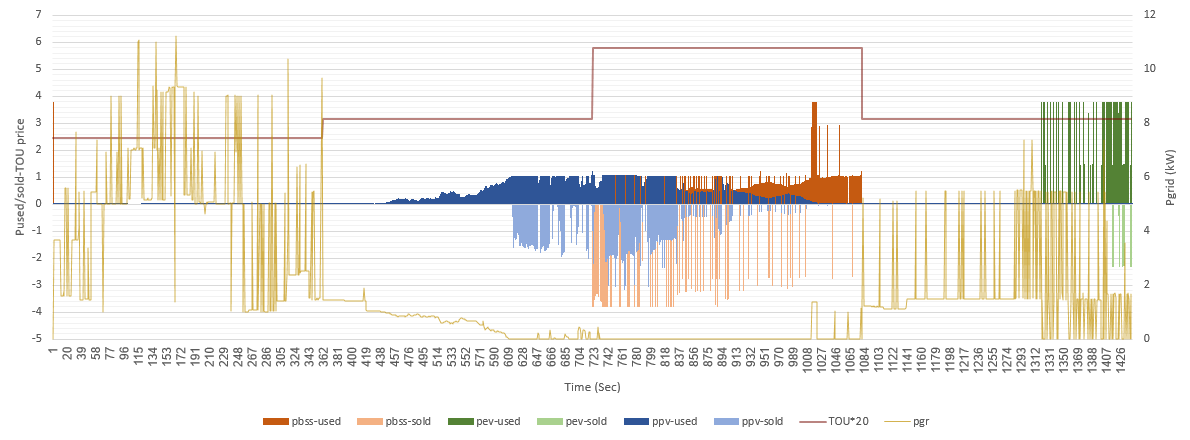}
    \caption{The amount of buying/selling from/to the grid, scenario I with Max-Median Deterministic Approach, TOU pricing.
.}
    \label{fig:buy_sell_scenario1_tou}
\end{figure}

\begin{figure}[H]
    \centering
    \includegraphics[width=\linewidth]{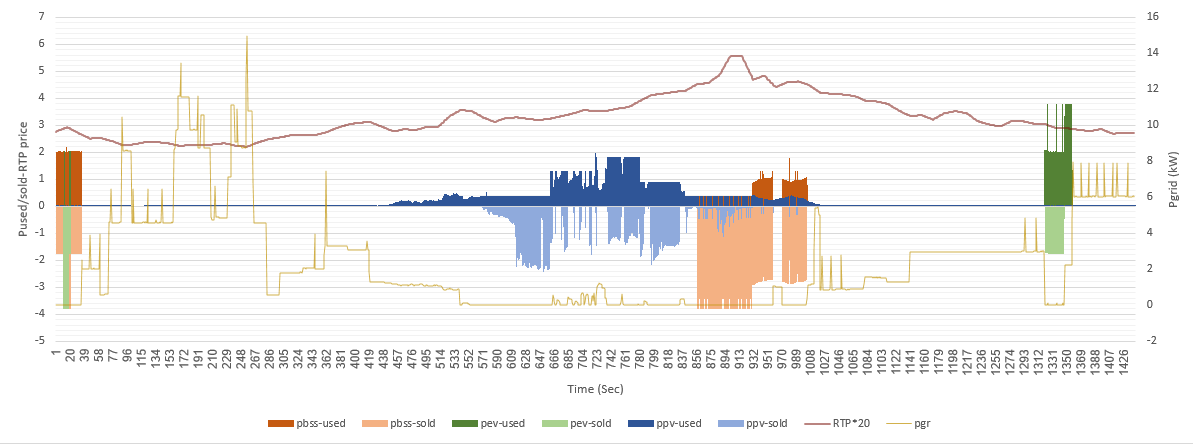}
    \caption{The amount of buying/selling from/to the grid, scenario I with Max-Avg Deterministic Approach, RTP.}
    \label{fig:buy_sell_scenario1_median_tou}
\end{figure}

\begin{figure}[H]
    \centering
    \includegraphics[width=\linewidth]{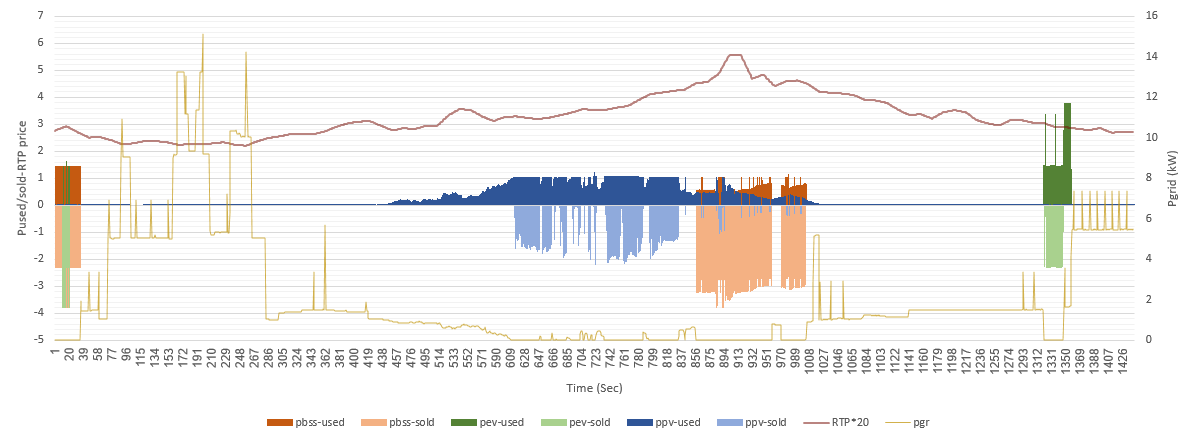}
    \caption{The amount of buying/selling from/to the grid, scenario I with Max-Median Deterministic Approach, RTP.}
    \label{fig:buy_sell_scenario1_rtp}
\end{figure}

\begin{figure}[H]
    \centering
    \includegraphics[width=\linewidth]{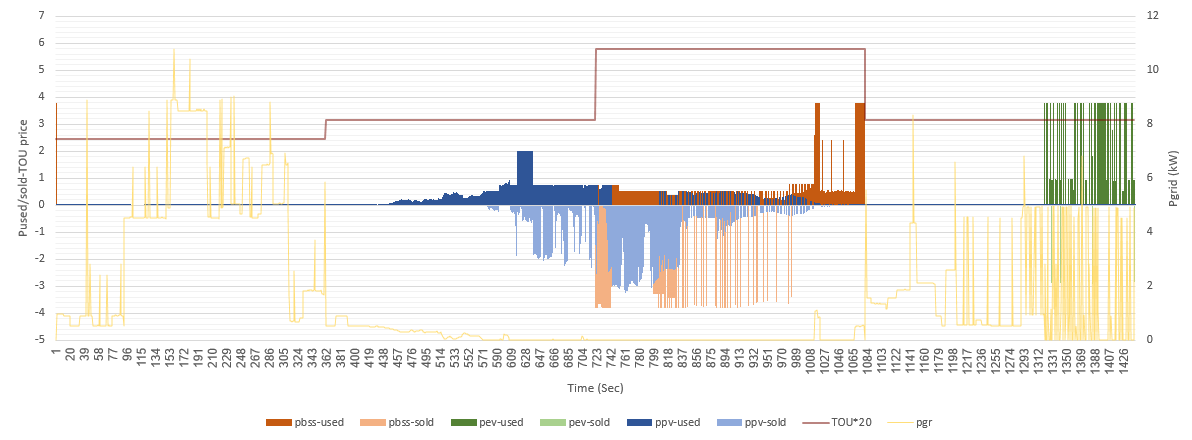}
    \caption{The amount of buying/selling from/to the grid, scenario II, TOU pricing.
.}
    \label{fig:buy_sell_scenario1_median_rtp}
\end{figure}

\begin{figure}[H]
    \centering
    \includegraphics[width=\linewidth]{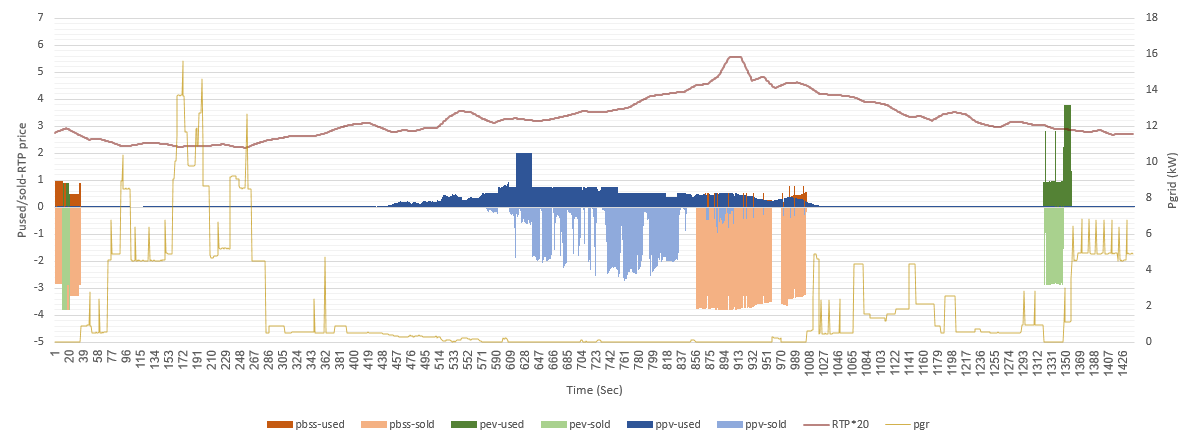}
    \caption{The amount of buying/selling from/to the grid, scenario II, RTP.
.}
    \label{fig:buy_sell_scenario2_tou}
\end{figure}

For a better comprehension of the impact of each of the management systems, Table \ref{tab:impact_table} has been prepared.

% \begin{table}[h]
%     \centering
%     % \captionsetup{font=small, labelfont=bf}
%     \caption{The impact of each management system on the total daily cost and net load fluctuations}
%     \label{tab:impact_table}
%     \renewcommand{\arraystretch}{1.2} % Adjusts row height for better readability
%     \setlength{\tabcolsep}{3pt} % Reduces column spacing to fit within margins
%     \resizebox{\columnwidth}{!}{ % Resizes table to fit within column width
%     \begin{tabular}{llcccccc}
%         \toprule
%         \textbf{Scenario} & \textbf{Approach} & \multicolumn{2}{c}{\textbf{PAR (\%)}} & \multicolumn{2}{c}{\textbf{SD (\%)}} & \multicolumn{2}{c}{\textbf{Daily Cost (\%)}} \\
%         \cmidrule(lr){3-4} \cmidrule(lr){5-6} \cmidrule(lr){7-8}
%         &  & TOU & RTP & TOU & RTP & TOU & RTP \\
%         \midrule
%         \multirow{}{}{I} & Max-Avg & \textbf{-44.19} & -27.30 & -15.46 & -6.01 & -8.27 & -22.19 \\
%         & Median-Avg & -31.67 & -17.30 & -18.17 & -3.22 & -28.04 & -2v7.26 \\
%         II & - & -20.89 & -0.95 & \textbf{-19.70} & -4.61 & \textbf{-62.05} & -57.28 \\
%         \bottomrule
%     \end{tabular}
%     } % End resizebox
% \end{table}
\begin{table}[h]
    \centering
    % \captionsetup{font=small, labelfont=bf}
    \caption{The impact of each management system on the total daily cost and net load fluctuations}
    \label{tab:impact_table}
    \renewcommand{\arraystretch}{1.2} % Adjusts row height for better readability
    \setlength{\tabcolsep}{3pt} % Reduces column spacing to fit within margins
    \resizebox{\columnwidth}{!}{ % Resizes table to fit within column width
    \begin{tabular}{llcccccc}
        \toprule
        \textbf{Scenario} & \textbf{Approach} & \multicolumn{2}{c}{\textbf{PAR (\%)}} & \multicolumn{2}{c}{\textbf{SD (\%)}} & \multicolumn{2}{c}{\textbf{Daily Cost (\%)}} \\
        \cmidrule(lr){3-4} \cmidrule(lr){5-6} \cmidrule(lr){7-8}
        &  & TOU & RTP & TOU & RTP & TOU & RTP \\
        \midrule
        \multirow{2}{*}{I} & Max-Avg   & \textbf{-44.19} & -27.30 & -15.46 & -6.01  & -8.27  & -22.19 \\
                           & Median-Avg & -31.67 & -17.30 & -18.17 & -3.22  & -28.04 & -27.26 \\
        II                & -          & -20.89  & -0.95  & \textbf{-19.70} & -4.61  & \textbf{-62.05} & -57.28 \\
        \bottomrule
    \end{tabular}
    } % End resizebox
\end{table}

\section{Conclusion}
\label{sec:5}
In this paper, an attempt was made to provide a comprehensive management system for the production and consumption of residential energy using the Internet of Things.
According to the load charts before and after planning, the proposed management system has succeeded in achieving the following:

\begin{itemize}
    \item Move controllable loads in such a way that consumption and charging are done when the price of electricity is lower, and discharge and selling are done at higher prices.
    \item Compensate for the momentary fluctuations caused by the integration of the solar system to an appropriate extent by predicting and planning using minutely recorded data.
    \item Satisfy the consumer by considering environmental indicators and conditions.
    \item Get more customer satisfaction with the help of very accurate forecasting, resulting in minimum changes in the household's desired consumption, reducing the total cost up to 62.05\%, fluctuations by up to 19.7\%, and smoothing the net power consumption profile up to 20.89\%.
\end{itemize}

In the studies carried out so far, the influence of non-schedulable loads in implementing the smart home planning and management system has been largely ignored. According to the diagrams achieved in this study, we notice the significant effect of the amount and time of consumption of uncontrollable loads in the intelligent planning of the controllable appliances of the residential house. Partial and dynamic changes in real-time consumption, in addition to cost savings, can essentially neutralize the fluctuations created on the net load caused by the integration of renewable generation with a high level of uncertainty and momentary fluctuations.

% In this regard, in order to satisfy consumers while solving energy management problems, it is necessary to have an accurate background of household consumption using appropriate forecasting models and techniques to increase forecasting accuracy. 
 As can be seen, by optimizing load scheduling (i.e. considering the probabilities of uncontrollable load) consumption and predicting these loads, we managed to reduce the total cost of consumption, fluctuations due to the integration of solar production, and load uncertainty. The use of suitable models and techniques to increase the accuracy of forecasting has resulted in the greatest cost reduction using both real-time and time-of-use pricing. In addition, the greatest decrease in standard deviation in time-of-use pricing is related to the use of predicted data, and the largest decrease in the peak-to-average ratio is achieved by using determined data in time-of-use pricing.

% \section*{References}
% \bibliographystyle{IEEEtran}
% % \bibliography{references}
\bibliographystyle{IEEEtran} 
\bibliography{references}
\end{document}